\documentclass[fleqn,usenatbib]{mnras}

\usepackage{newtxtext,newtxmath,ulem}

\usepackage[T1]{fontenc}

\DeclareRobustCommand{\VAN}[3]{#2}
\let\VANthebibliography\thebibliography
\def\thebibliography{\DeclareRobustCommand{\VAN}[3]{##3}\VANthebibliography}

\usepackage{graphicx}	
\usepackage{amsmath}	
 
\usepackage{amssymb}	

\newcommand{\ac}{anticentre}
\newcommand{\kms}{\,km\,s$^{-1}$}

\title[Pristine XIII: The very metal-poor thin disc]{The Pristine survey XIII: Uncovering the very metal-poor tail of the thin disc.}

\author[Emma Fern\'andez-Alvar et al.]{
Emma Fern\'andez-Alvar,$^{1}$\thanks{E-mail: emmafalvar@gmail.com}
Georges Kordopatis,$^{1}$
Vanessa Hill,$^{1}$
Else Starkenburg,$^{2}$
Akshara Viswanathan,$^{2}$
\newauthor
Nicolas F. Martin,$^{3}$
Guillaume F. Thomas,$^{4,5}$
Julio F. Navarro,$^{6}$
Khyati Malhan,$^{7}$
Federico Sestito,$^{6}$
\newauthor
Jonay I. Gonz\'alez Hern\'andez,$^{4,5}$
and Raymond G. Carlberg$^{8}$
\\
$^{1}$Universit\'e C\^ote d'Azur, Observatoire de la C\^ote d'Azur, CNRS, Laboratoire Lagrange, Bd de l'Observatoire, CS 34229, 06304 Nice cedex 4, France \\
$^{2}$Kapteyn Astronomical Institute, University of Groningen, PO Box 800, NL-9700 AV Groningen, the Netherlands\\
$^{3}$Universit \'e de Strasbourg, CNRS, Observatoire astronomique de Strasbourg, UMR 7550, F-67000 Strasbourg, France\\
$^{4}$Instituto de Astrof\'isica de Canarias, E-38205 La Laguna, Tenerife, Spain \\
$^{5}$Universidad de La Laguna, Dpto. Astrof\'isica, E-38206 La Laguna, Tenerife, Spain \\
$^{6}$Department of Physics and Astronomy, University of Victoria, Victoria, BC, Canada V8P 5C2\\
$^{7}$The  Oskar  Klein  Centre  for  Cosmoparticle  Physics,  Department  of Physics,  Stockholm  University,  AlbaNova,  10691  Stockholm,  Sweden \\
$^{8}$Department of Astronomy \& Astrophysics, University of Toronto, Toronto, ON M5S 3H4, Canada
}

\date{Accepted XXX. Received YYY; in original form ZZZ}

\pubyear{2021}

\begin{document}
\label{firstpage}
\pagerange{\pageref{firstpage}--\pageref{lastpage}}
\maketitle

\begin{abstract}
We evaluate the rotational velocity of stars observed by the Pristine survey towards the Galactic \ac\, spanning a wide range of metallicities from the extremely metal-poor regime ($\mathrm{[Fe/H]}<-3$\,) to nearly solar metallicity.
In the Galactic \ac\ direction, the rotational velocity ($V_{\phi}$) is similar to the tangential velocity in the galactic longitude direction ($V_{\ell}$). This allows us to estimate $V_{\phi}$ from Gaia early data-release 3 (Gaia EDR3) proper motions for stars without radial velocity measurements. This substantially increases the sample of stars in the outer disc with estimated rotational velocities.
Our stellar sample towards the \ac\ is dominated by a kinematical thin disc with a mean rotation of $\sim -220$ \kms. However, our analysis reveals the presence of more stellar substructures. The most intriguing is a well populated extension of the kinematical thin disc  down to $\mathrm{[Fe/H]} \sim -2$\,. A scarser fast rotating population reaching the extremely metal-poor regime, down to $\mathrm{[Fe/H]} \sim -3.5$\, is also detected, but without statistical significance to unambiguously state whether this is the extremely metal-poor extension of the thin disc or the high rotating tail of hotter structures (like the thick disc or the halo). In addition, a more slowly rotating kinematical thick disc component is also required to explain the observed $V_{\ell}$ distribution at $\mathrm{[Fe/H]} > -1.5$\,. Furthermore, we detect signatures of a ``heated disc'', the so-called \textit{Splash}, at metallicities higher than $\sim-1.5$. Finally, at $\mathrm{[Fe/H]} < -1.5$\, our \ac\ sample is dominated by a kinematical halo with a net prograde motion.
\end{abstract}




\section{Introduction}

Our understanding of the processes that formed the Galaxy benefits from information encoded in the properties of stars, mainly their chemical composition and their orbital motion. In particular, the analysis of the first stellar populations formed in the Universe contribute valuable hints to the earliest stages of galactic formation \citep{fb02}. These stars are characterized by their very low metal content and have been the object of thorough searches over the last decades -- see the reviews \citet{bc05} and \citet{frebel15}.


To unveil which processes led to the current configuration of the Galaxy we need the orbital characterization of its stars. The velocity of a star with respect to the Sun is measured in three components: the radial velocity ($\mathrm{v}_{\mathrm{rad}}$) and the proper motions ($\mu_{\alpha}$ and $\mu_{\delta}$), which correspond to the velocity projections along the line-of-sight, the right ascension and the declination directions, respectively. For convenience, sometimes the proper motions are expressed in the galactic reference system ($\mu_{\ell}$ and $\mu_{b}$). The vast improvement in terms of accuracy, precision and coverage of astrometric data (proper motions and parallaxes) provided by the Gaia mission \citep{gaia2016, gaia18b, gaia2020} has revealed the dynamical configuration of the stellar Galactic populations with an unprecedented level of detail. As some examples, we have now an expanded insight into the degree at which the Galaxy is out of equilibrium \citep{gaia18dyn}, evidence that the Galactic halo in the inner regions is dominated by an accreted system and a heated protodisc \citep{helmi18, haywood18, myeong18, gallart19, dimatteo19, myeong19, belokurov20}.



The Gaia mission is also planning to provide spectroscopic metallicities -- from the CaII near-infrared triplet region, as explained in \citet{recioblanco16}. However, no chemical abundances estimates are published yet, radial velocities are only available for the brighter part of the catalogue, and even from ground-based complement, samples of low-metallicity stars with full kinematics are still of modest size. The Pristine survey \citep{stark17} was conceived as a narrow-band photometric survey centered in the CaII doublet H\&K at 3933 and 3968 \AA, a spectral feature very sensitive to the metallicity\footnote{From now on we will refer to metallicity as the abundance of iron with respect to hydrogen, and we will express it as the difference with respect to the Sun abundance in logarithmic scale: $\mathrm{[Fe/H]} = \log_{10} (\mathrm{N_{Fe}/N_{H}})_{\mathrm{star}} - \log_{10} (\mathrm{N_{Fe}/N_{H}})_{\odot}$} content of the star when combined with broad band photometry. Its main goal is to provide a comprehensive view of the stellar metallicity distribution of the metal-poor Galaxy, with a particular focus on discovering the pristine metallicity stars born from the products of the first episodes of star formation. The survey, carried out with the 3.6-meter optical/infrared Canadian-French-Hawaiian Telescope (CFHT) on the Mauna Kea Hawaiian Observatory, already covers over 5000 square degrees on the sky, observing millions of targets over most of the northern Galactic hemisphere. From their photometry, a metallicity estimate is derived for every object observed. Thus, the combination of the Pristine and the Gaia surveys opens huge possibilities for the characterization of the first stellar populations and their link with the subsequent star formation in the Galaxy. 

Indeed, kinematical analyses of metal-poor stars ([Fe/H] < -1) making use of Gaia astrometry have already provided new results regarding the characterization of the first generations of stars. For instance, \citet{sestito19} discovered extremely and ultra metal-poor stars ([Fe/H] < -3 and [Fe/H] < -4) rotating fast in prograde orbits close to the Galactic plane, never detected before. Di Matteo et al. (2019) showed signature of rotation in stars down to [Fe/H] < -2. Carollo et al. (2019) obtained evidence that the thick disk comprises two distinct and overlapping stellar populations with different kinematic properties and chemical compositions. More studies after them have confirmed these results.


Unfortunately, radial velocities can only be measured from stellar spectra \footnote{However, see \citet{ld21}. }, and only $\sim$0.5\% of stars in the Pristine survey have a spectroscopic follow-up \citep{caffau17, youakim17, bonifacio19, monelli19, venn20, caffau20, kielty20}. Consequently, the fraction of Pristine stars with full velocity characterization (proper motions and radial velocity measurements) is extremely low, and it is worth searching for specific directions on the sky where geometry cancels the contribution of the radial velocity in some of the velocity components with respect to the Galactic center.


A number of studies have already made use of this kind of approach in the past. For instance, before the advent of the Gaia survey, there were few accurate proper motions measurements, but a number of publications used privileged directions where the stellar rotational velocity, $V_{\phi}$, is identical to $\mathrm{v}_{\mathrm{rad}}$ in order to evaluate the rotation of the Galaxy (e.g., \citealt{morrison90, wyse16, kordopatis13, kordopatis17}). 

With the advent of Gaia, the problem has now  become the opposite: we have high-quality proper motions but the lack of spectra prevents us from having a $\mathrm{v}_{\mathrm{rad}}$ estimation for the majority of stars in the survey. At exactly the \ac\ direction $V_{\phi}$ is perpendicular to the line-of-sight and latitude directions and, consequently, the projection of $V_{\phi}$ along these directions is null. In other words, to derive $V_{\phi}$ from the observed $\mathrm{v}_{\mathrm{rad}}$, $\mu_{b}$ and $\mu_{\ell}$ velocity components only the latter is required. This fact allows us to study the $V_{\phi}$ distributions from proper motions only, without radial velocities. This is still a good approximation up to a few degrees away from the \ac\, as we will demonstrate in this paper. 

Some previous works have already inferred the rotational velocity of stars from proper motions only towards the \ac. For example, \citet{thomas19} explored the disc flaring with this approximation, using blue stragglers. \citet{anbeers2020,anbeers21} showed how stars rotating slower than the disc, i.e., the kinematical halo, is structured in several components based on their location on the $V_{\phi}$ vs. [Fe/H] space. They detected signatures of the \textit{Splash} thick disc \citep{belokurov20}, the metal-weak thick disc \citet{norris85}, the Gaia-Enceladus-Sausage -- GES \citep{belokurov18, helmi18} accreted system, and an additional component dominated by stars on retrograde orbits. In addition, as part of the science verification of the Gaia early third data release (Gaia EDR3), \citet{antoja21} revealed the complex structure of the disc towards the \ac\ with, among others, the detection of velocity oscillations throughout the plane, and the identification of stellar overdensities associated to Monoceros, TriAnd, and the \ac\ stream. Further structures have been detected recently in \citet{laporte21} also using this approximation.

Stars in the outer disc of the Milky-Way are less affected by the Galactic potential and the dynamical relaxing timescales are longer, thus favouring the detection of kinematical signatures of the processes that the Galaxy underwent (e.g., \citealt{binney08}). The Pristine survey provides a map of stars in a wide range of metallicities, with uniquely robust estimates at the lowest metallicities when compared to other photometric metallicity inferences \citep{stark17, youakim17, aguado19}, giving key information about star formation from the earliest stages of galactic formation. For all these reasons, the rotational velocity distribution of the stars presented in this work provides a unique insight on this regard. 

In this work we aim to combine the Pristine metallicities with Gaia astrometry to explore the rotational velocity distribution of stars in the Galaxy towards the \ac\ direction with a special emphasis on metal-poor stars. Indeed, Pristine, with its robust photometric metallicity estimates in the very (VMP, $\mathrm{[Fe/H]} < -2$\,) and extremely (EMP, $\mathrm{[Fe/H]} < -3$\,) metal-poor regime, provides a unique opportunity to probe the kinematics of these populations in the \ac\ direction. We explain our data selection, the methodology to determine the rotational velocity from proper motions only, and the adopted stellar metallicities in Section 2. Section 3 explores the rotational velocity distributions as a function of [Fe/H]. We discuss the implications of our results in Section 4. Finally, our conclusions are summarized in Section 5.

\section{Methodology and data selection}

We cross-correlate Pristine observations, obtained between 2015 and 2020, with Gaia EDR3 \citep{gaiaedr321} and the Sloan Digital Sky Survey (SDSS) photometry \citep{abaz09, fuku96, gunn98, doi10} in order to estimate [Fe/H] and the rotational velocity component. In order to justify the selection criteria applied, we will firstly explain the bases of our methodology.

\subsection{Azimuthal velocities derived from the proper motions towards the \ac.}
\label{sec:vl_computation}

\begin{figure}
\centering
\includegraphics[scale=0.5, trim= 15 0 0 0]{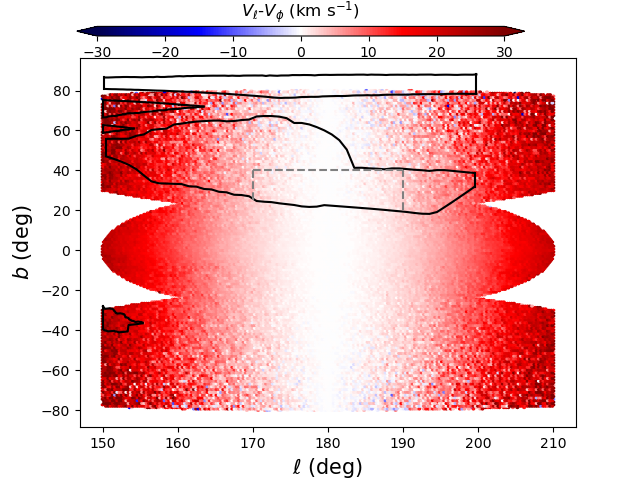}  
\caption{Average $V_{\phi}$ - $V_{\ell}$  as a function of galactic coordinates ($\ell$, $b$) measured in the Gaia Mock catalogue. The plot is a superposition of two projected cones centered at ($\ell$,$b$) = (180,0)\,\degr, ($\ell$,$b$) = (0,-50)\,\degr and ($\ell$,$b$) = (0,50)\,\degr with a radius of 30\,\degr\,. The grey dashed delimits the selected region for our analysis.}
\label{simul}
\end{figure}

\begin{figure}
\centering
\includegraphics[scale=0.3, trim=0 0 -100 0]{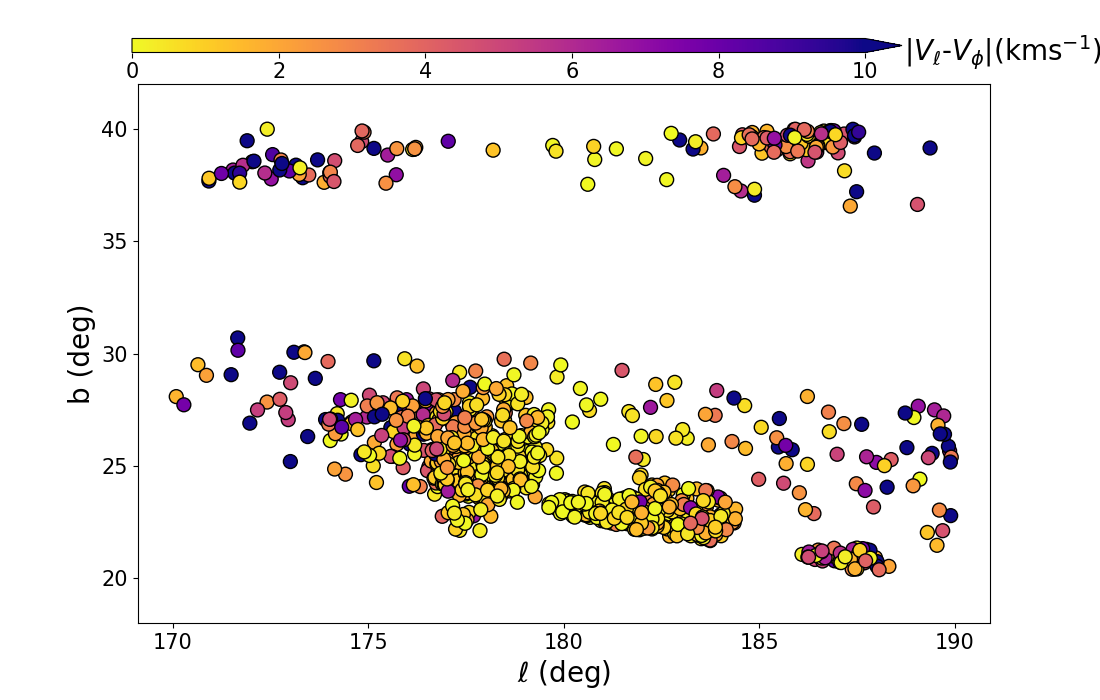} 
\includegraphics[scale=0.3, trim=0 0 -100 0]{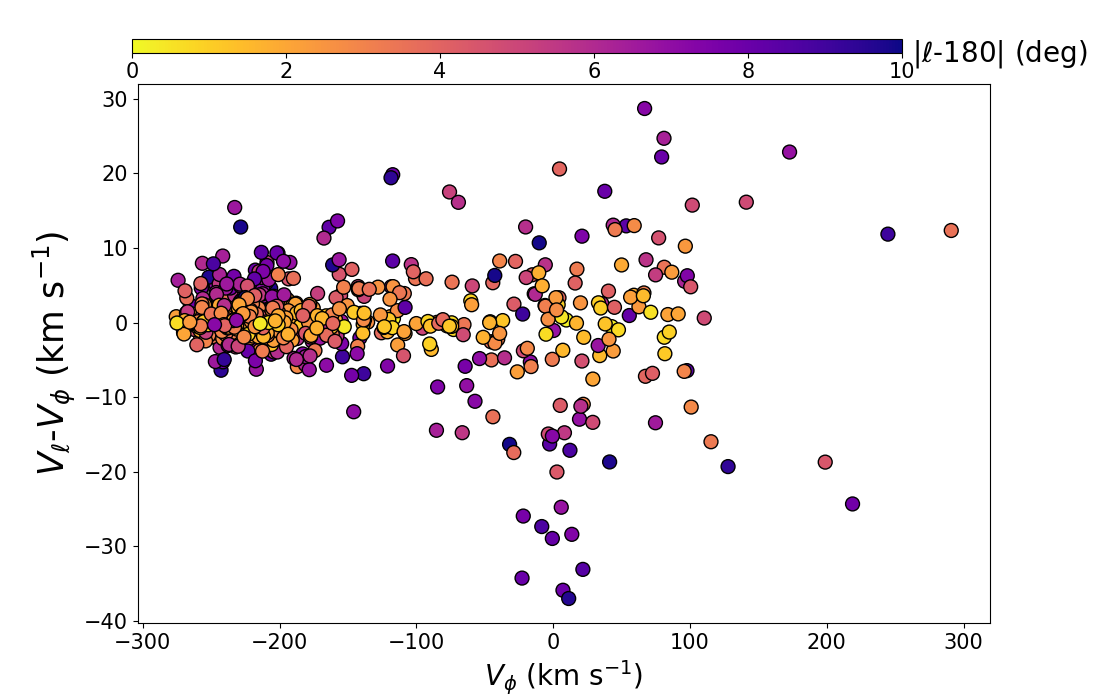} 
\caption{Top panel: $\ell$ and $b$ distribution of our training sample located in the $ac$ colour coded with the difference of $V_{\ell}$ compared with $V_{\phi}$ measured from Gaia EDR3 proper motions and the spectroscopic $\mathrm{v}_{\mathrm{rad}}$. Bottom panel: The same $V_{\ell}$ - $V_{\phi}$ differences as shown in the top panel but now as a function of the $V_{\phi}$, colour coded by |$\ell-180$\degr|.}
\label{vphicomp_training}
\end{figure}

\begin{figure}
\centering
\includegraphics[scale=0.35, trim= 0 50 0 30]{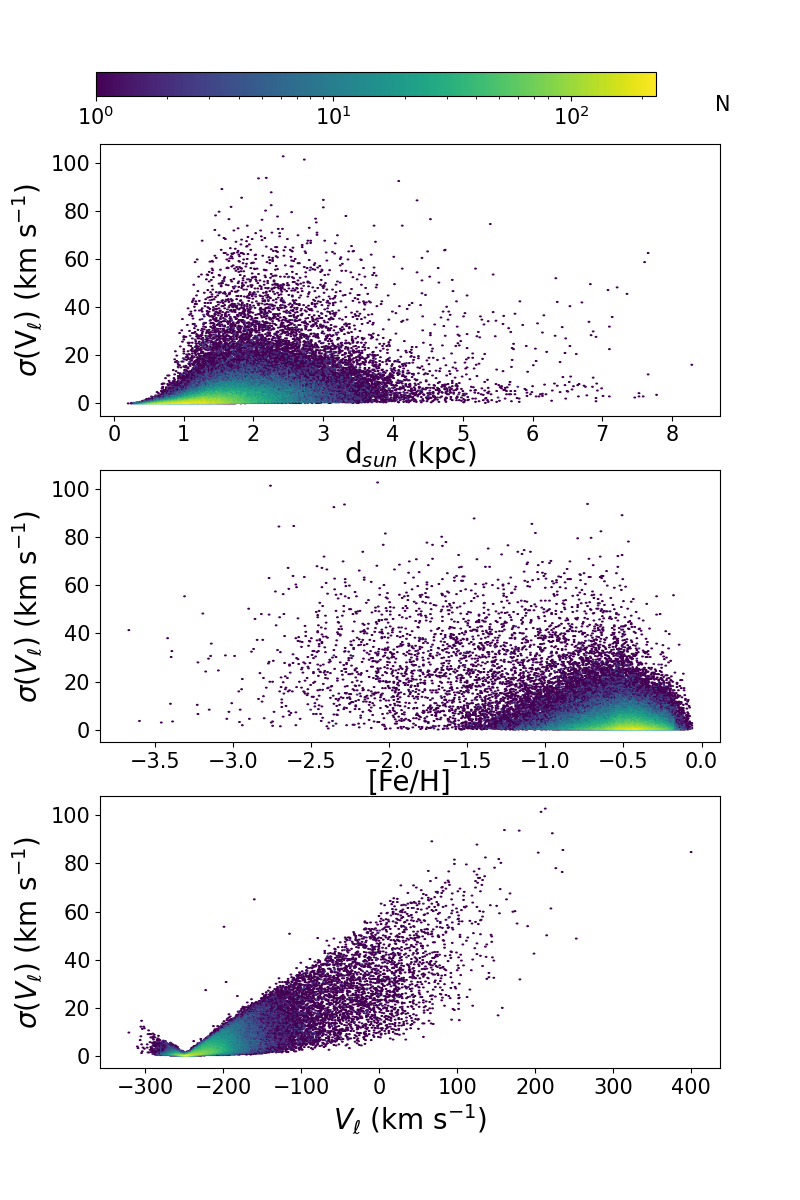} 
\caption{$V_{\ell}$ uncertainties as a function of the line-of-sight distance, $\mathrm{d}_{sun}$, [Fe/H] and $V_{\ell}$. The uncertainties are computed as the standard deviation of 1000 $V_{\ell}$ Monte-Carlo realisations calculated by drawing from normal distributions of the parallax and proper motions uncertainties.}
\label{MCVl}
\end{figure}


First, we present the geometrical relations between proper motions and the rotational velocity component. Proper motions (in mas $\mathrm{yr}^{-1}$) are transformed from the equatorial to the galactic coordinate system ($\ell$, $b$) through the equations in \citet{poleski13}, that perform the rotations needed to change reference system:

\begin{eqnarray}
\mu_{\ell} = \frac{1}{\cos(b)(c_1\mu_{\alpha} + c_2\mu_{\delta})} \\
\mu_b = \frac{1}{\cos(b)(c_2\mu_{\alpha} + c_1\mu_{\delta})} ,
\end{eqnarray}

with $c_1$ and $c_2$ defined as

\begin{eqnarray}
c_1 & = & \sin(\delta_{GP})\cos(\delta) - \cos(\delta_{GP})\sin(\delta)\cos(\alpha - \alpha_{GP})  \\
c_2 & = & \cos(\delta_{GP})\sin(\alpha - \alpha_{GP}) 
\end{eqnarray}

and ($\alpha_{GP}$, $\delta_{GP}$) = (192.860, 27.128) degrees are the equatorial coordinates of the North Galactic Pole \citep{ESA97}.


Then we transform from angular to linear velocities:
 
\begin{eqnarray}
V_{\ell} & = &  4.74 \mathrm{d}_{sun} \mu_{\ell} \\
V_b & = & 4.74 \mathrm{d}_{sun} \mu_b  
\end{eqnarray}
where $\mathrm{d}_{sun}$ is the distance in kpc inferred from the inverse of the Gaia parallaxes\footnote{In this paper, we consider only stars with a low relative uncertainty in parallaxes < 20\%. As a consequence, the subsequent uncertainties in distances are also low and there is no need to apply priors to obtain realistic distances -- see \citet{bailerjones15}.}.
$V_{\ell}$ and $V_b$ are the linear projections of the proper motions over the galactic coordinates ($\ell$, $b$). 

We adopt the right-handed galactocentric cartesian and cylindrical coordinate systems, i.e., the X axis positive towards the Sun, the Z axis positive towards the North Galactic pole, and the rotation of the Local Standard of Rest (LSR) negative. In this system, then, a negative value of $V_{\phi}$ corresponds to a prograde rotation. We also adopt the galactocentric distance of the Sun $R_{sun} = 8.3$\,kpc \citep{schonrich12}. 

The cartesian velocity components U, V and W with respect to the LSR are linked to $V_{\ell}$ and $V_b$ through the following equations:

\begin{eqnarray}
U & = & \mathrm{v}_{\mathrm{rad}}\cos(\ell)\cos(b) - V_\ell \sin(\ell) - V_b \cos(\ell) \sin(b) \\
V & = & \mathrm{v}_{\mathrm{rad}}\sin(\ell)\cos(b) + V_\ell \cos(\ell) - V_b \sin(\ell) \sin(b) \\
W & = & \mathrm{v}_{\mathrm{rad}}\sin(b) + V_b \cos(b) \label{ww}
\end{eqnarray}


and we correct them for the solar motion with respect to the LSR ($U_{\odot}$, $V_{\odot}$, $W_{\odot}$) = (11.1,12.24,7.25) \kms \citep{schonrich10} and the rotational velocity component of the LSR $V_{\mathrm{LSR}} = -238$\,\kms~ \citep{schonrich12}.

$U, V$ and $W$ can be transformed to the cylindrical coordinate system defined by the coordinates R,$\phi$ and z by:

\begin{eqnarray}
V_R & = & U \cos(\phi) + V \sin(\phi) \\
V_{\phi} & = & -U \sin(\phi) + V \cos(\phi) \\
V_z & = & W \label{vz}
\end{eqnarray}
where $V_R$, $V_{\phi}$ and $V_z$ are the radial, rotational and vertical velocity components. Then, $V_{\phi}$ can be expressed as a function of ($\mathrm{v}_{\mathrm{rad}}$,$V_{\ell}$, $V_{b}$) by:

\begin{equation}
\begin{aligned}
V_{\phi} = {} & \mathrm{v}_{\mathrm{rad}}[\sin(\ell) \cos(b) \cos(\phi) - \cos(\ell) \cos(b) \sin(\phi)] \\
& + V_\ell[ \sin(\ell) \sin(\phi) + \cos(\ell) \cos(\phi)] \\ 
& + V_b[ \cos(\ell) \sin(b) \sin(\phi)   - \sin(\ell) \sin(b) \cos(\phi)]
\end{aligned}
\end{equation}


Thus, for a star located in the direction of the \ac\ where $\ell$ = 180\degr, $b$ = 0\degr and $\phi$ = 0\degr, the rotational velocity component of the star, $V_{\phi}$, is identical to the linear projection of the proper motion measured over the direction of the galactic longitude, $V_{\ell}$. 

In the case of stars observed away from the \ac\, there is a dependence of $V_{\phi}$ on $\mathrm{v}_{\mathrm{rad}}$ and $V_{b}$ which is not zero. However, assimilating $V_{\phi}$ to $V_{\ell}$ remains a good aproximation in the surroundings of the \ac. Considering the sky area observed by Pristine towards the \ac\, (marked with black contours in Figure \ref{simul}), we evaluated which are the regions where we can use $V_{\ell}$ as a good aproximation of $V_{\phi}$ for our Pristine stars. We evaluated the Gaia eDR3 mock catalogue provided by the Gaia Archive\footnote{We made use of the \texttt{gaiaedr3.gaia\_source\_simulation} catalogue available through \url{https://gea.esac.esa.int/archive/}.} in the directions ($\ell$,$b$) = (180,0)\,\degr, ($\ell$,$b$) = (0,-50)\,\degr and ($\ell$,$b$) = (0,50)\,\degr with a radius of 30\,\degr. We calculated $V_{\phi}$ following Equations 7 to 12, and we measured the differences between $V_{\phi}$ and $V_{\ell}$ (obtained based on Equations 1 to 6), as shown with the color scale in Figure \ref{simul}. This analysis shows that $V_{\ell}$ follows $V_{\phi}$ with an accuracy better than 10 \kms~ in ($170$\degr$ < \ell < 190$\degr). Figure \ref{simul} shows that at latitudes higher than $b = 40$\degr and lower than $b = -40$\degr the deviation increases more rapidly when departing from $\ell = 180$\degr than at $-40 < b < 40$\degr. 

Looking at the Pristine footprint we decided to constrain our analysis to the region ($170$\degr$ < \ell < 190$\degr, $20 < b < 40$\degr), to select a monolithic region in ($\ell$,$b$) with low $|V_{\ell}$ - $V_{\phi}|$ deviations to keep a simpler relation with the distance to the plane and the center, z and R.

As an additional verification, we computed the $V_{\phi}$ for the Pristine stars in the \ac\ that were also observed spectroscopically (the spectroscopic sample used for the metallicity calibration that we describe in \ref{met_det}) and had a measured $\mathrm{v}_{\mathrm{rad}}$ from previous works (see \citealt{yanny09} and \citealt{aguado19} for more details), and compared their $V_{\phi}$ with the resulting $V_{\ell}$ from proper motions. We evaluated the stars located in the selected region ($170$\degr$ < \ell < 190$\degr, $20 < b < 40$\degr) and applied the same astrometry and photometry quality criteria as for the main analysis that we explain in Section \ref{met_det}. The comparison of $V_{\ell}$ with $V_{\phi}$, shown in Figure \ref{vphicomp_training}, reveals as expected that the scatter in $V_{\ell}$ - $V_{\phi}$ increases with $|\ell-180|$ (i.e., $\ell$ departs from the \ac). For the bulk of stars the differences are within the expected uncertainty from the simulations, i.e., lower than 10 km $\mathrm{s}^{-1}$, with differences as large as 40 \kms in some cases. The stars showing largest differences are those with $V_{\phi}$ close to 0 or positive, i.e., those stars with no rotation or moving on retrograde orbits. These stars are expected to have a high radial velocity component. For such stars located away from the \ac\ their $V_{\ell}$ is large because of the contribution of the projection of the radial velocity component, explaining the discrepancy with their low or opposite sign $V_{\phi}$.



We thus show that it is possible to characterize the rotational motion of stars without radial velocity measurements in the \ac\ direction within the area ($170$\degr$ < \ell < 190$\degr, $-40$\degr$ < b < 40$\degr). This gives us the opportunity to combine the huge Gaia EDR3 database of proper motions with the Pristine targets, which, in the \ac\, corresponds to a sample of 414,292 stars. We correct parallaxes from the zero point obtained from quasars, $-0.017$\,mas \citep{lind2020a} and remove those stars with a relative uncertainty in parallaxes higher than 20\%, and parallaxes equal to, or lower than 0.

We measured the typical uncertainties in $V_{\ell}$ via 1000 Monte-Carlo realisations drawing from normal distributions of the uncertainties in parallax and proper motions. The mean of the derived uncertainties in $V_{\ell}$ is 4 \kms. The parameter that dominates the $V_{\ell}$ uncertainty is the error in the parallax. Figure \ref{MCVl} shows the dependence of the uncertainties with respect to the line-of-sight distance, [Fe/H] and $V_{\ell}$ itself. Figure \ref{MCVl} also shows that for some stars, in particular those with larger $V_{\ell}$, the uncertainties are significantly much larger than 4 \kms. We decided to include them and verify in the subsequent analysis that such large uncertainties do not affect our results.

\subsection{The [Fe/H] determination.}
\label{met_det}

We follow the same methodology as described in \citet{stark17} -- hereafter S17 -- to convert from the Pristine narrow-band Ca\,II-HK photometric observations to metallicity estimates. The Ca\,II-HK doublet at 3933 and 3968 \AA\ is a very prominent spectral feature very sensitive to stellar metallicity variations, in particular in metal-poor stars with $\mathrm{[Fe/H]} < -1$\,. The Ca\,II-HK doublet is also temperature sensitive, thus, a metallicity estimation can be performed by combining the CaHK magnitude with broad-band colours, in a colour-colour space that separates out the temperature and metallicity effects (see right panel on figure 3 in S17). The position of a star in this space can then be transformed into a metallicity estimate. 
In brief, we calibrate metallicity estimates using a stellar sample (that we will call training sample), composed of stars with spectroscopic metallicities, complemented by colours computed from synthetic spectra at the most metal-poor end. In the Pristine plus broad-band colour-colour plane, we define "pixels" for which we derive a mean metallicity based on all the training sample stars that are located in this colour-colour pixel. Thus, every location in the photometric space has a metallicity value associated to it, and a Pristine target, based on its Ca\,II-HK and SDSS photometry alone, is then assigned a metallicity based on this map. For a more detailed explanation we refer the reader to S17.

The spectroscopic sample used on this calibration (training sample) has evolved since S17, and now include metallicities inferred from the SEGUE Spectroscopic Parameters Pipeline (SSPP) and released as part of the SEGUE DR12 \citep{yanny09, alam15} of all Pristine targets with a SEGUE counterpart, and also metallicity estimates of very metal-poor stars ($\mathrm{[Fe/H]} \lesssim -2$\,) from the Pristine collaboration follow-up spectroscopy \citep{aguado19}, complemented with a relation derived from synthetic spectra at the most metal-poor end (as described in S17). 

The resulting calibration provides reliable photometric metallicity estimates for metal-poor FGK stars \citep[S17]{aguado19}.  An upper metallicity cut at [Fe/H] = $-0.2$ is applied to the results, because of the high errors and biases found for those. Indeed, for metal-rich stars ([Fe/H] $> -1$), the Ca\,II-HK lines are heavily saturated, and depend heavily on the stars luminosity (or gravity). In the following, we show how we improve the situation in the intermediate metallicity range  ([Fe/H]$\sim -1$), by taking the stellar luminosity into account in the calibration.




\subsubsection{Refining the Pristine metallicity calibration with Dwarf/Giant classification}
\label{feh_comparison}

\begin{figure}
\centering
\includegraphics[scale=0.4]{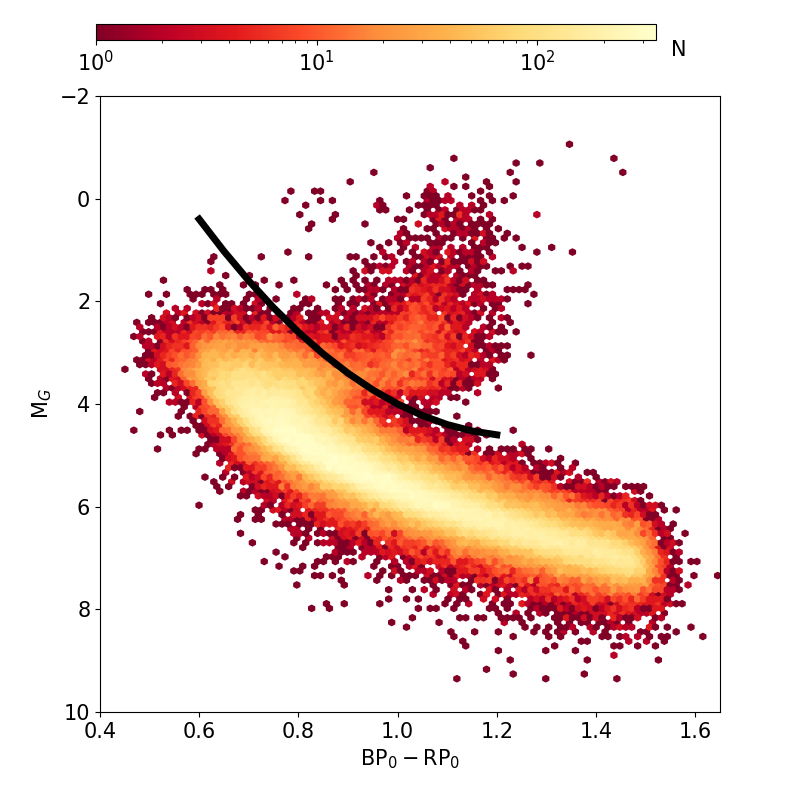}  
\caption{Colour-magnitude diagram of our Pristine sample towards the \ac\, based on Gaia EDR3 photometry. The black line indicates the boundary chosen to classify these stars in dwarf and giants.}
\label{CMD}
\end{figure}

\begin{figure}
\centering
\includegraphics[scale=0.4]{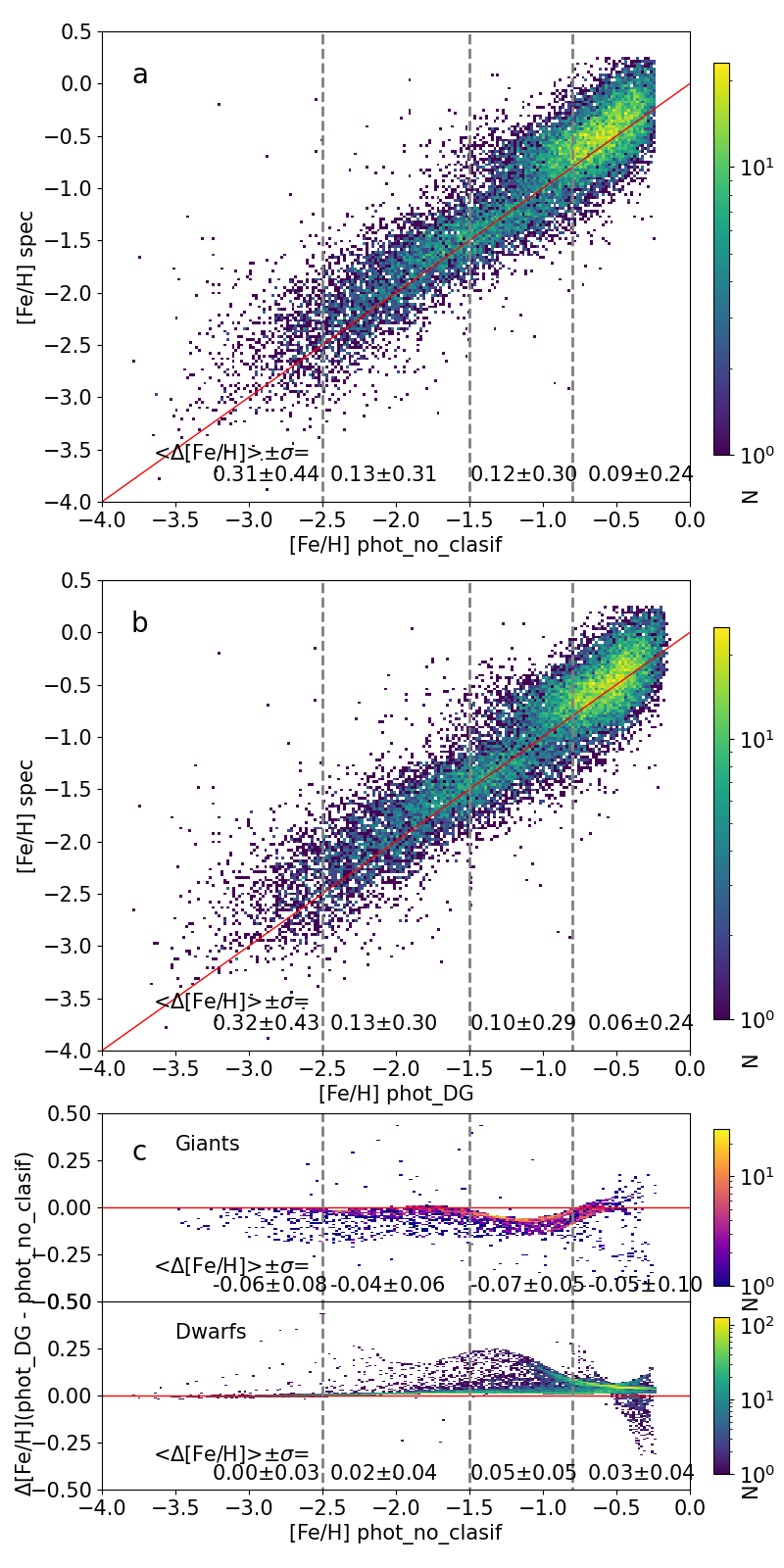}    
\caption{2-d histograms showing the comparison between the [Fe/H] inferred spectroscopically (by the SDSS stellar parameters pipeline) and the ones inferred photometrically. Top panel (a): the adopted photometric [Fe/H] is the one based on the global calibration. Medium panel (b): the adopted photometric [Fe/H] takes into account the dwarf/giant classification. The two bottom panels (c) show the differences between both photometric calibrations, separately for giants and dwarfs}. Mean differences and dispersions ($\rm \Delta \mathrm{[Fe/H]} \pm \sigma$) are given for various metallicity intervals (defined by the vertical dashed lines). 
\label{comp_feh_mean}
\end{figure}

\begin{figure*}
\centering
\includegraphics[scale=0.4]{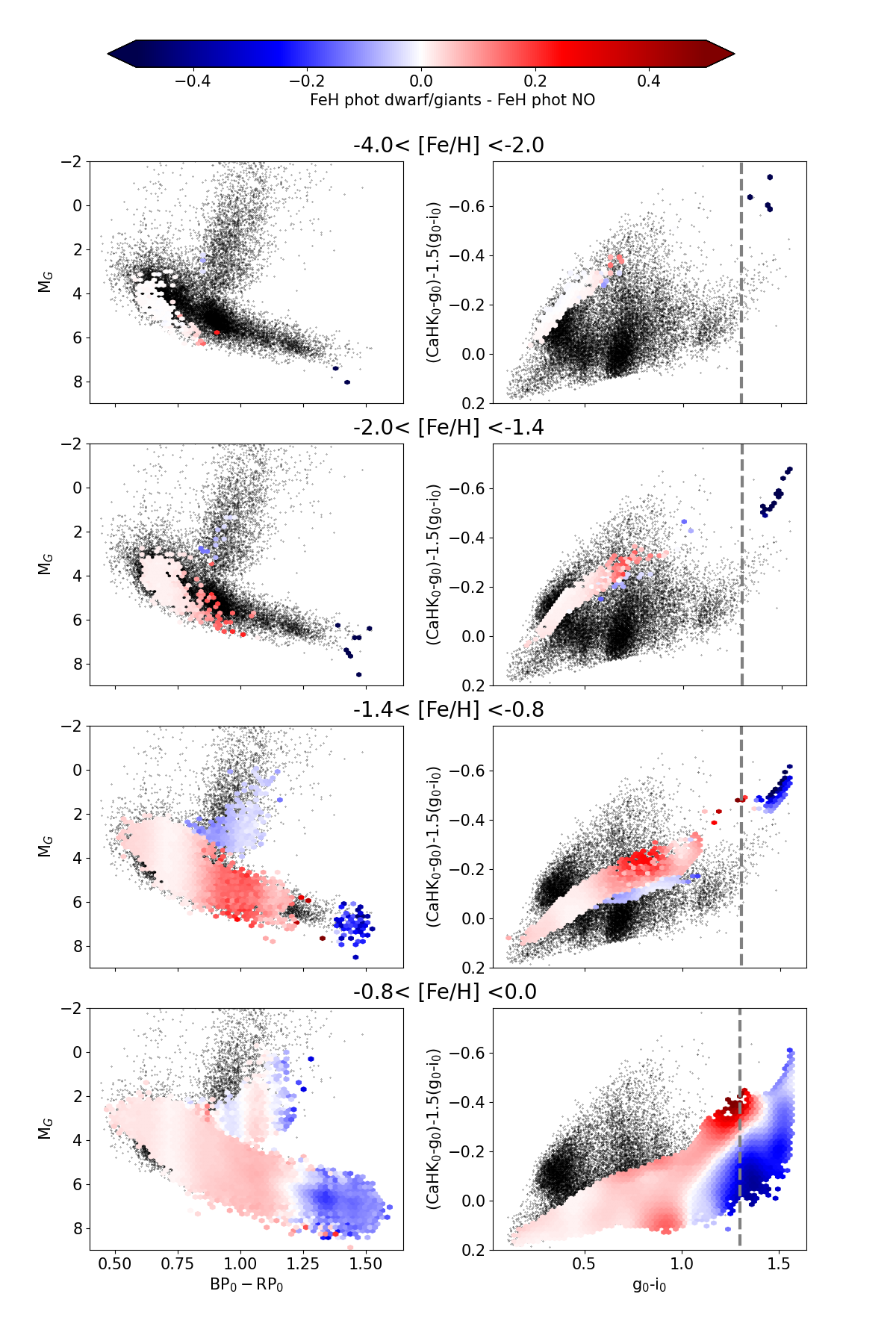} 
\caption{Left panels: colour magnitude diagram based on {\it Gaia} EDR3 photometry. The background black dots correspond to the training sample. The mean differences (calculated in hexagonal bins on the CMD) of the [Fe/H] estimates resulting from the calibration with the dwarf/giant classification and the global one for our \ac\ Pristine sample (split in metallicity bins) are overplotted. Right panels: Same as left panels but in the colour-colour space (combining Pristine and SDSS photometry) used for the [Fe/H] calibration.}
\label{CMD_diffeh}
\end{figure*}

At relatively high metallicities, $\mathrm{[Fe/H]} > -1$\,, a position in the Pristine colour-colour space is not solely defined by the metallicity and effective temperature (traced by
the broad band $(g-i)_{0}$ colour), but also by the stellar surface gravity, making the calibration uncertain, unless the gravity is also known accurately, and taken into account in the calibration (thus also requiring a training sample with accurate gravities). 

In this work we improve the adopted [Fe/H], making use of the {\it Gaia} absolute $M_{\rm G}$ magnitude, and $BP-RP$ colour in order to classify our stars as dwarfs or giants. We estimate the colour excess of each target due to dust reddening based on the Schlegel maps \citep{schlegel98}. These maps provide the colour excess corresponding to the B-V Johnson \& Cousins magnitude system based on their location on the sky. We transform them to the {\it Gaia} magnitude system through the following equations:

\begin{eqnarray}
G_0 = G - 0.86117*3.1*E(B-V) \\
BP_0 = BP - 1.06126*3.1*E(B-V) \\
RP_0 = RP - 0.64753*3.1*E(B-V) 
\end{eqnarray}
where $G_0$, $BP_0$ and $RP_0$ are the de-reddened $G$, $BP$ and $RP$ Gaia apparent magnitudes, and the coefficients multiplying $E(B-V)$ are obtained through the estimator in \url{http://stev.oapd.inaf.it/cgi-bin/cmd_3.4} based on the relations of \citet{cardelli89} and \citet{odonnell94}. We use the SDSS de-reddened magnitudes and the Pristine de-reddening is described in S17. 

We perform two different metallicity calibrations, each following the same principle as in S17: the first uses a training sample containing only dwarf stars, while the second is based on a training sample containing only giants, classified based on their $M_{G}$ vs. $\mathrm{(}BP-RP\mathrm{)}_{0}$ CMD diagram, from Gaia EDR3. Then we assign to each star the [Fe/H] derived from the corresponding metallicity calibration, depending on their locus in the same {\it Gaia} CMD shown in Figure \ref{CMD}. The internal metallicity uncertainties computed for our \ac\ sample are of 0.1 in average, with a dispersion of 0.04 for dwarfs, and 0.1 with a sigma of 0.02 for giants.


In Figure \ref{comp_feh_mean} (panel c) we show the overall difference between the calibration with and without taking into account the luminosities of the training sample stars, while in Figure \ref{CMD_diffeh}, we show the location of the training sample and our \ac\ sample in a colour-magnitude diagram and in the Pristine colour-colour space in order to 
verify the impact of our dwarf/giant metallicity estimation. As expected, in the colour range where the stellar gravities are similar on the main sequence and giant branch, there is little difference between the two calibrations (eg. base of the RGB, subgiants, and main sequence around G stars). In fact, in most of the parameter space, the difference between the global and dwarf/giants metallicity calibrations is smaller or equal to 0.2, and as expected, the differences are most important in the metal-rich regime. There are, however, special areas where the differences are larger. In particular, for the reddest colours, the metallicity estimates for the global and dwarf/giants metallicity calibrations can reach up to 0.4 and even 1.0 . While it is indeed at the reddest colours that the stellar gravities differ the most between a giant and a dwarf star, and hence where one expects the largest impact on the metallicity estimate, we note that the training sample exhibits a severy lack of metal-poor main sequence stars (K dwarfs). As it is clearly visible in both the CMD, and, most prominently, in the Pristine colour-colour space, our \ac\ sample extends to the regions where metal-poor K dwarfs are expected to lie, yet, with hardly any training sample, casting doubts on the metallicities estimated for these stars. We therefore excluded from our sample stars with $(g-i)_{0}$ higher than 1.3. We stress that this cut further discriminates against metal-rich stars. 

The comparison of these photometric metallicities with spectroscopic [Fe/H], displayed in Figure \ref{comp_feh_mean} (panel a and b). The overall performance of the calibration with and without taking into account the luminosity of stars is rather similar, with a very slight decrease of the overall bias and dispersion of the retrieved photometric metallicities with our new calibration separating stars according to their luminosities. The overall behaviour is preserved, with a bias against the most metal-rich stars because of the cut imposed by the calibration at [Fe/H]$>-0.2$, and a slight underestimation of photometric metallicities at the lowest metallicities ($\mathrm{[Fe/H]} < -2$), where the mean offset and dispersion of the differences are the highest, $0.32 \pm 0.43$. One should however bear in mind that spectroscopic measurements for these very metal-poor stars are also less precise, and certainly concur to inflate the dispersion in that regime.

There are however subtle differences between the calibrations with and without taking the stellar luminosity into account, that will play a role in the present investigation. In particular, around spectroscopic metallicities $\mathrm{[Fe/H]_{phot}}$ around $-0.7$, a regime where the spectroscopic metallicities are the most reliable, a number of stars deviate from the one-to-one relation and show photometric metallicities in the range $\mathrm{[Fe/H]_{phot}}= -0.8$ to $-1.5$ in the original calibration. This population of deviating stars is reduced with our new calibration. Panel c of Figure \ref{comp_feh_mean} shows that this is the metallicity domain where the two calibrations differ most, both for giants where the new calibration gives slightly lower metallicities, and and for dwarfs where the new calibration gives higher metallicities. The dwarf stars in the training sample thus contribute to reduce the bias and dispersion observed between spectroscopic and photometric metallicities with our new calibration in the range $\mathrm{[Fe/H]_{phot}}= -0.8$ to $-1.5$. Since our anticenter sample comprises mostly dwarf stars (the ratio of dwarfs to giants is higher than in the training sample, see Figure \ref{CMD}), we expect this new calibration to be even better in that problematic metallicity range. There however remains a potential contamination of metal-rich stars when selecting low-metallicity targets based on photometric estimations that should be taken into consideration when interpreting our results. We will address the impact of such contamination in subsection \ref{contamination}.

\subsubsection{Additional quality selection cuts.}

There are several caveats regarding the metallicity calibration based on this approach, which could potentially affect [Fe/H] estimates for some of our targets. For example, we detect a group of white dwarfs among our \ac\ sample although this stellar type is not included in our training sample. White dwarfs lack Ca absorption lines and therefore they end up with an extremely low metallicity estimate if we misinterpret them as main sequence or giant stars. We filter them out by removing targets with $M_{\rm G}$ < 10. In \citet{youakim20} it was shown that young stars are another case that may be prone to erroneous [Fe/H] estimate. Following \citet{youakim20} we  reject targets with $(g-i)_{0} < 0.6$, $(u-g)_{0} > 1.15$ and $(u-g)_{0} > 1.5(g-i)_{0}+0.6$. We finally reject stars in an unphysical location in the colour-colour space, as defined by the synthetic spectra computed with no metals (i.e., above the black curve in the left panel of Figure 3 of S17).

As explained before, Pristine metallicities partially rely on broad-band SDSS photometry obtained several years before the actual Pristine survey (which started in 2016). Therefore, any variability in the stellar flux may result in a biased derived metallicity. For that reason, we clean our sample of possible variable stars. 

We use the standard deviation $\sigma \mathrm{flux}$ of the G-band fluxes (which have been obtained from 2014 to 2017) as an indicator of their variability. It is provided by the Gaia EDR3 archive\footnote{\url{https://gea.esac.esa.int/archive/}} as the 
\texttt{phot\_g\_mean\_flux\_over\_error}
parameter, normalized to the square root of the number of observations contributing to the G mean flux, $\sqrt{\texttt{phot\_g\_n\_obs}}$ : 

\begin{equation}
\sigma \mathrm{flux} = \frac{ \sqrt{\texttt{phot\_g\_n\_obs} }}{\texttt{phot\_g\_mean\_flux\_over\_error}}
\end{equation}

We remove objects verifying the following empirically defined relation:

\begin{equation}
\sigma \mathrm{flux} > \frac{(\sigma \mathrm{flux} - 16)^2}{800} * \mathrm{max}(\sigma \mathrm{flux} - 15, 0) + 0.015.
\end{equation}


In order to avoid binary systems that could lead to wrong astrometric estimates we use the Renormalised Unit Weight Error (RUWE) parameter provided by the Gaia archive, keeping only targets with a RUWE parameter lower than 1.4, as suggested by the Gaia Collaboration \citep{lind2020b}.

Our final sample after applying all the selection cuts comprises 152,079 stars.



\section{Results}

\begin{figure}
\centering
\includegraphics[scale=0.36]{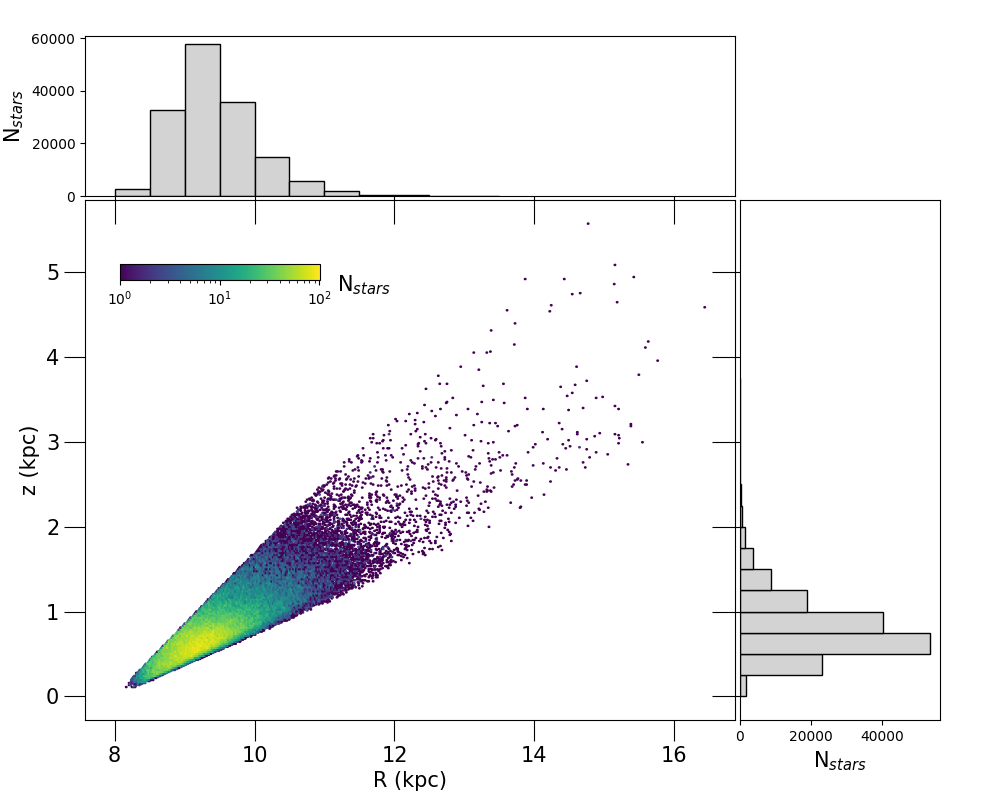}  
\caption{Distribution of the hight above the plane, z, vs. galactocentric radius, R, for our Pristine \ac\ stellar sample, colour coded by the number of stars.
Histograms at the top and at the right correspond to the R and z distributions, respectively.}
\label{Rz}
\end{figure}

\begin{figure*}
\centering
\includegraphics[scale=0.5]{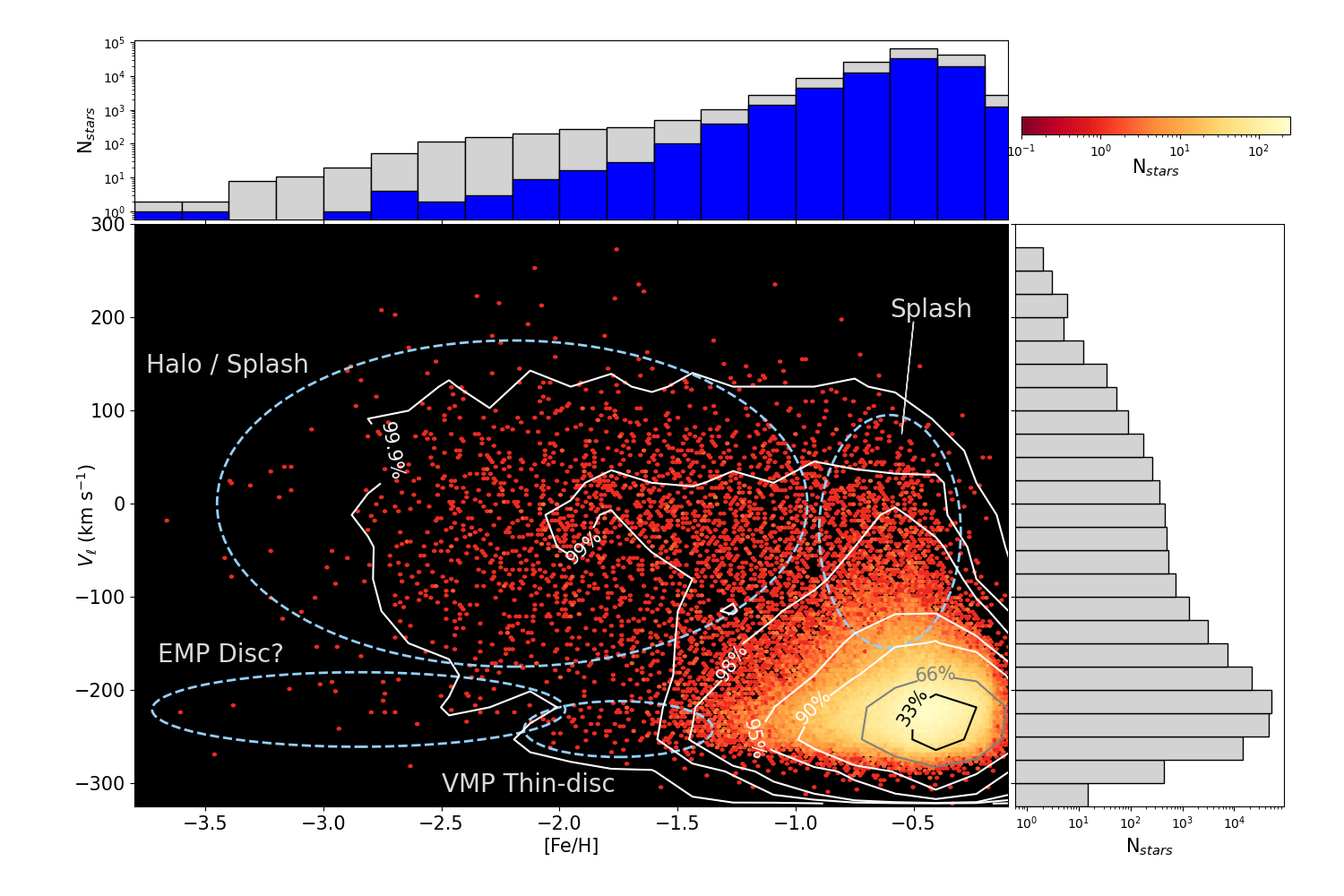}  

\caption{$V_{\ell}$ as a function of [Fe/H], colour coded by density. The density is measured in bins of 10 \kms and 0.1 dex. Contour lines show the 33, 66, 98, 99 and 99.9\% of the cumulative distribution. The most evident stellar substructures are pointed out with annotated blue ellipses. These are: a presumed extremely metal-poor disc (EMP disc), the very metal-poor thin disc (VMP thin disc), the Splash, and the halo (with the likely contribution of metal-poor Splash stars). Stars on prograde motion are those with $V_{\ell}$ < 0, while stars on retrograde motion have $V_{\ell}$ > 0. Top panel: Metallicity distribution function of all the \ac\ sample in light grey and of the fast rotators ($V_{\ell} < -220$\kms) in blue. Right panel: $V_{\ell}$ distribution of all the \ac\ sample. Both distributions are plotted in logarithmic scale. Note: the color code lower limit is set to 0.1 due to color contrast purposes.}
\label{chachi}
\end{figure*}


\begin{figure*}
\centering
\includegraphics[scale=0.4]{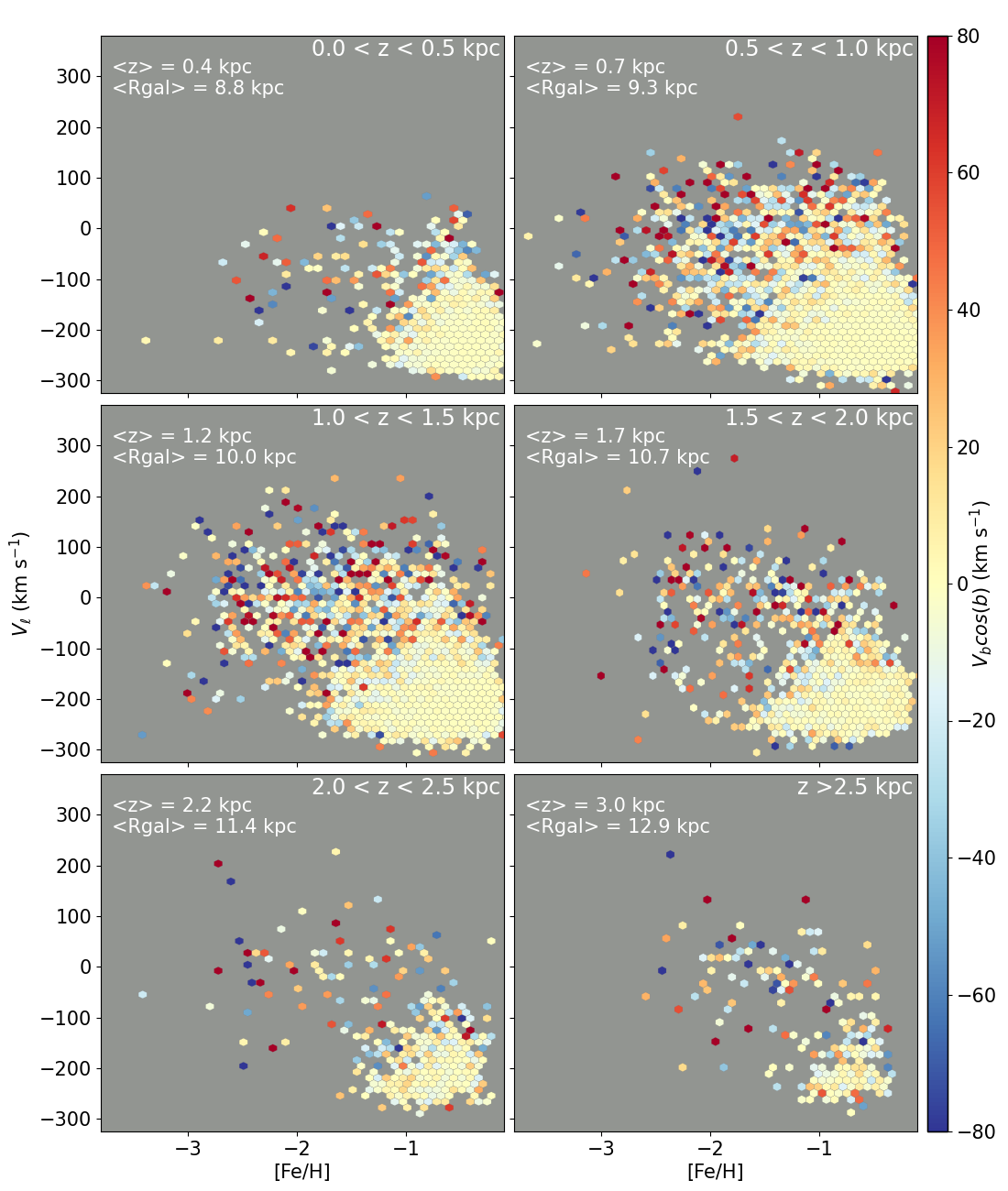}  
\caption{$V_{\ell}$ as a function of [Fe/H], colour coded by $V_{b}\cos(b)$, in bins of z. The colour code is computed as the mean $V_{b}\cos(b)$ in hexagonal bins, and the mean height above the plane $<\mathrm{z}>$ and galactocentric radius $<\mathrm{Rgal}>$ are given in each panel.}
\label{Vb}
\end{figure*}

\begin{figure*}
\centering
\includegraphics[scale=0.6]{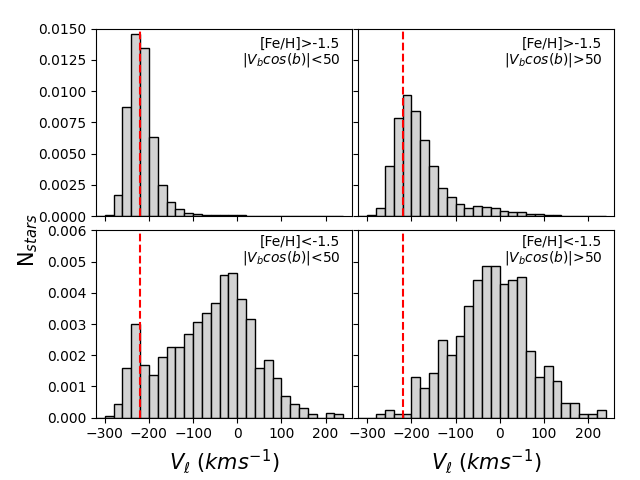}  
\caption{$V_{\ell}$ distribution for stars with $\mathrm{[Fe/H]} > -1.5$\, (top panels) and with $\mathrm{[Fe/H]} < -1.5$\, (bottom panels), and with low vertical velocities |$V_{b}\cos(b)$| < 50\,\kms (left panels) and high vertical velocities |$V_{b}\cos(b)$| > 50\,\kms (right panels). Velocities lower than $V_{\ell} = -220$\kms (the red dashed line) are those characteristic of thin disc stars.}
\label{Vb_histos}
\end{figure*}

\begin{figure}
\centering
\includegraphics[scale=0.5]{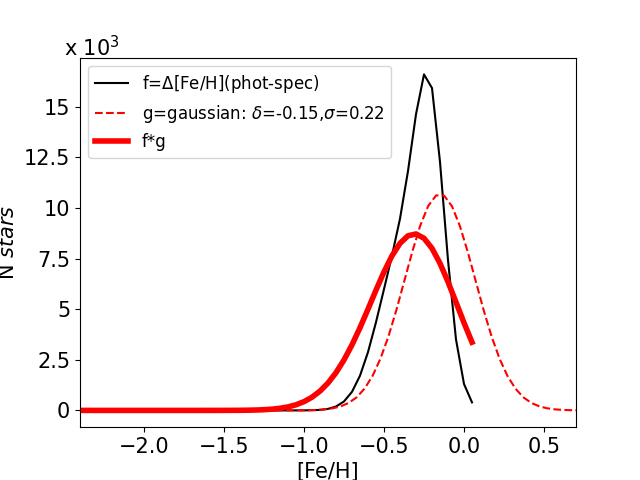}  
\caption{This figure illustrates the metallicities that Pristine is expected to recover (red continuous line) if we feed the survey with solely thin disc targets distributed according to a Gaussian MDF centered at $-0.15$ with a sigma of $0.22$ (red dashed line). The transformation is obtained by convolving this MDF with an error function (in black), derived from targets in the range $\mathrm{[Fe/H]}_{\rm spec}=[-0.2,0]$\, of Figure \ref{comp_feh_mean}. }
\label{model_cont}
\end{figure}

\begin{figure*}
\centering
\includegraphics[scale=0.4]{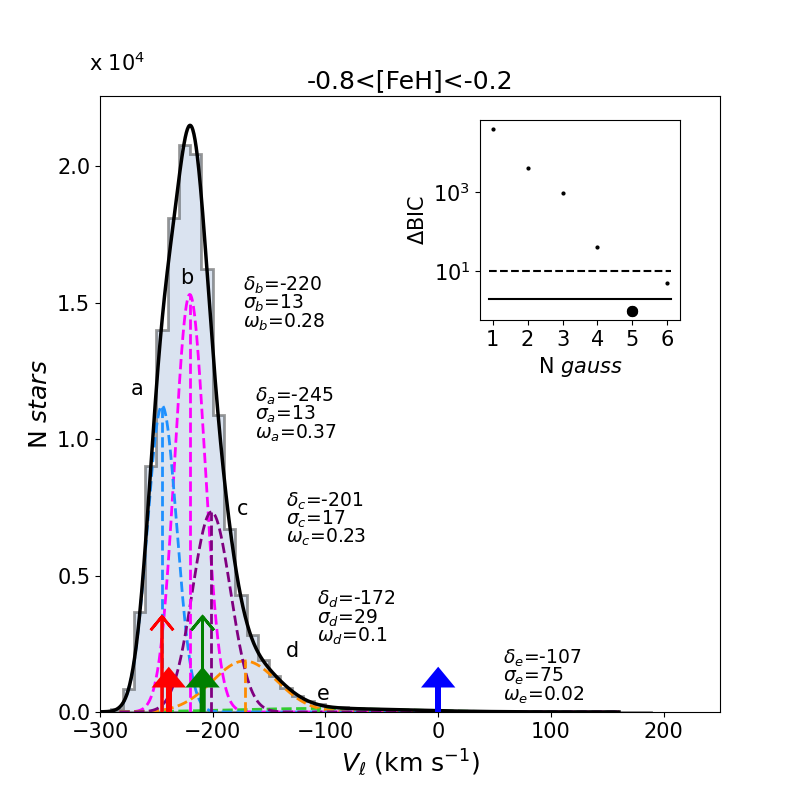}  
\includegraphics[scale=0.4]{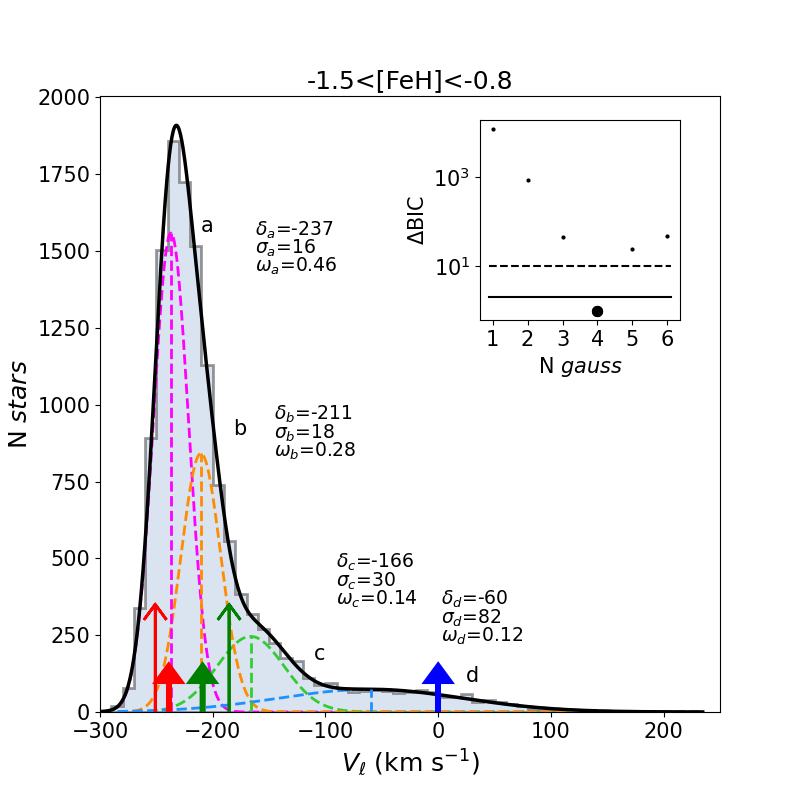}  
\caption{Best Gaussian Mixture Model fits of the $V_{\ell}$ values for stars with $-0.8 < \mathrm{[Fe/H]} < -0.2$\, (left panel) and $-1.5 < \mathrm{[Fe/H]} < -0.8$\, (right panel) are plotted with continuous black lines. The corresponding individual gaussians are in colored dashed lines, together with their means, widths and relative weights annotated next to them. At the top right corner we show the differences between the BIC of each model and the one with the best fit, as well as the reference values of statistical significance for BIC differences of 10 (dashed black line) and 2 (continuous black line).} 
\label{histos_mr}
\end{figure*}

\begin{figure*}
\centering
\includegraphics[scale=0.4, trim= 0 -240 0 0]{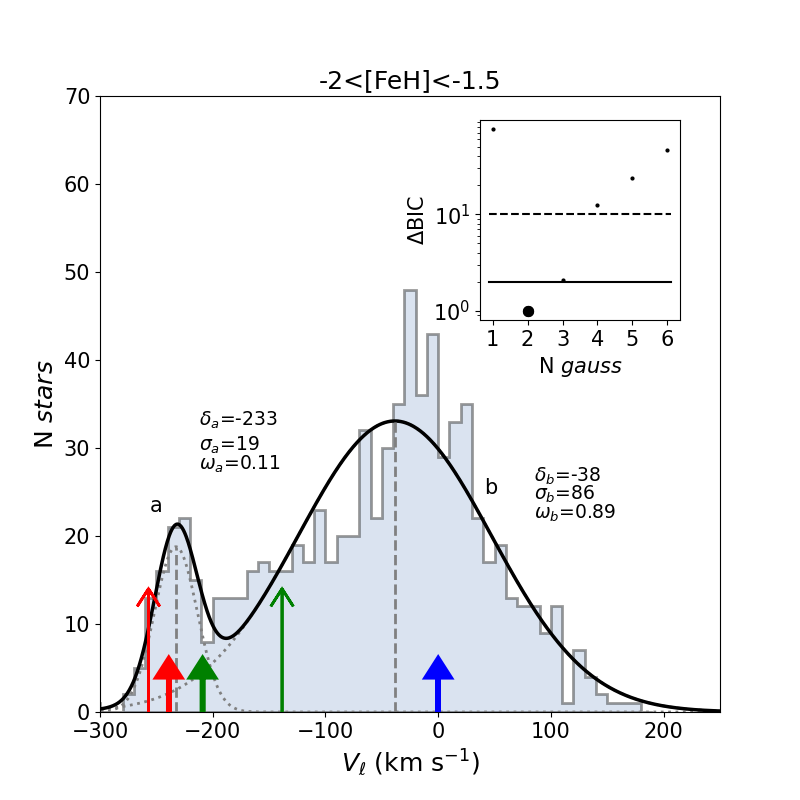}  
\includegraphics[scale=0.35]{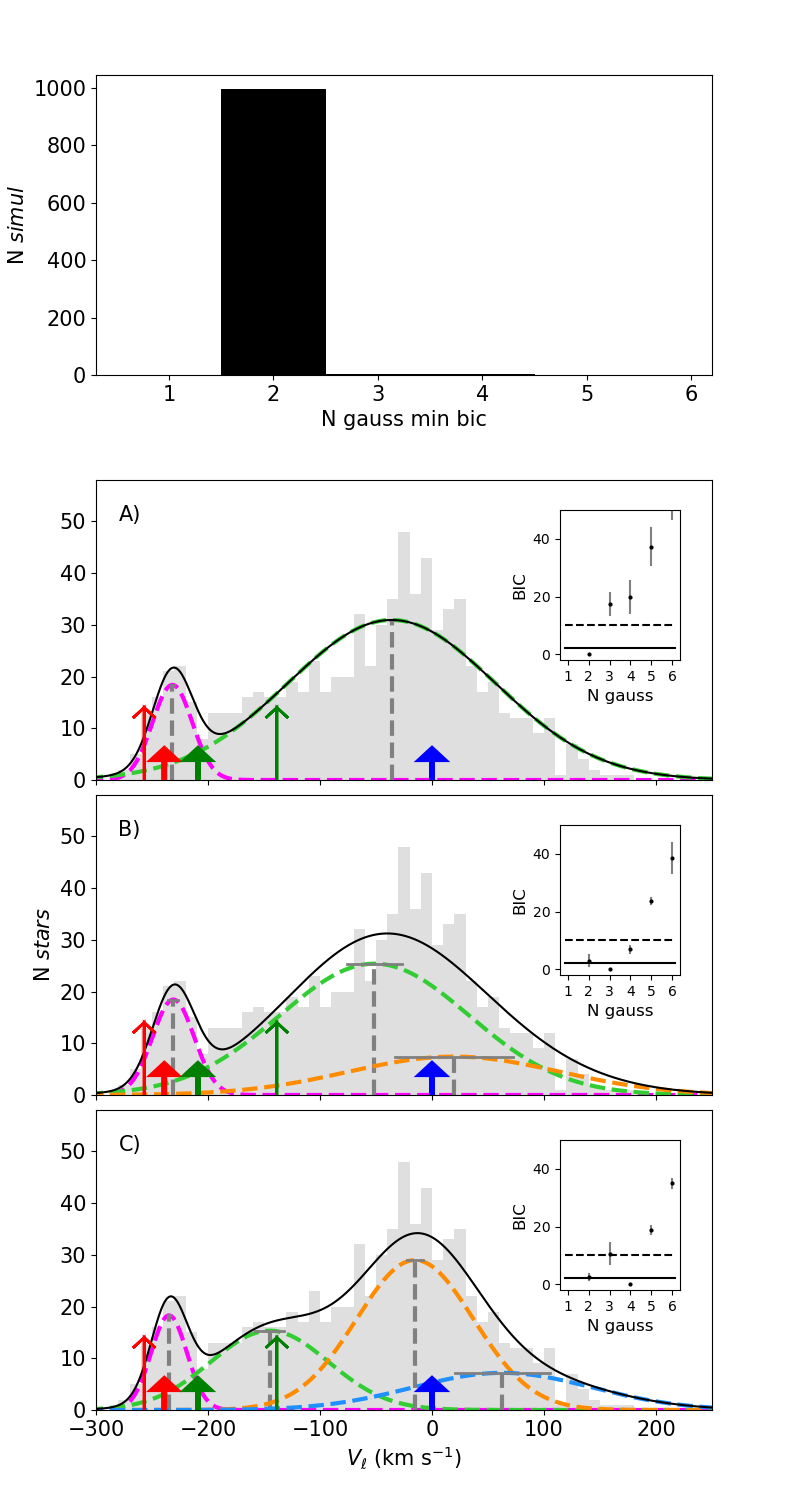}  
\caption{Left panel: same as Figure \ref{histos_mr} but for the metallicity range $-2 < \mathrm{[Fe/H]} < -1.5$\,. Right panel: top plot shows  the distribution of the number of components with the best fit obtained from 1000 Monte-Carlo realisations; the bottom panels show the mean gaussians when this best fit was a two-gaussian model (panel A), three-gaussian model (panel B) or four-gaussian model (panel C). Horizontal lines above the gaussians correspond to uncertainty on the centers of the distributions.
 On the right top corner of these panels the mean difference of the BIC between each model and the best fit for the 1000 Monte-Carlo realisations are plotted, as well as the standard deviations as error bars. The reference values of statistical statistical significance for BIC differences of 2 (dashed black line) and 10 (continuous black line) are also indicated.}
\label{m2}
\end{figure*}

\begin{figure*}
\centering
\includegraphics[scale=0.4, trim= 0 -240 0 0]{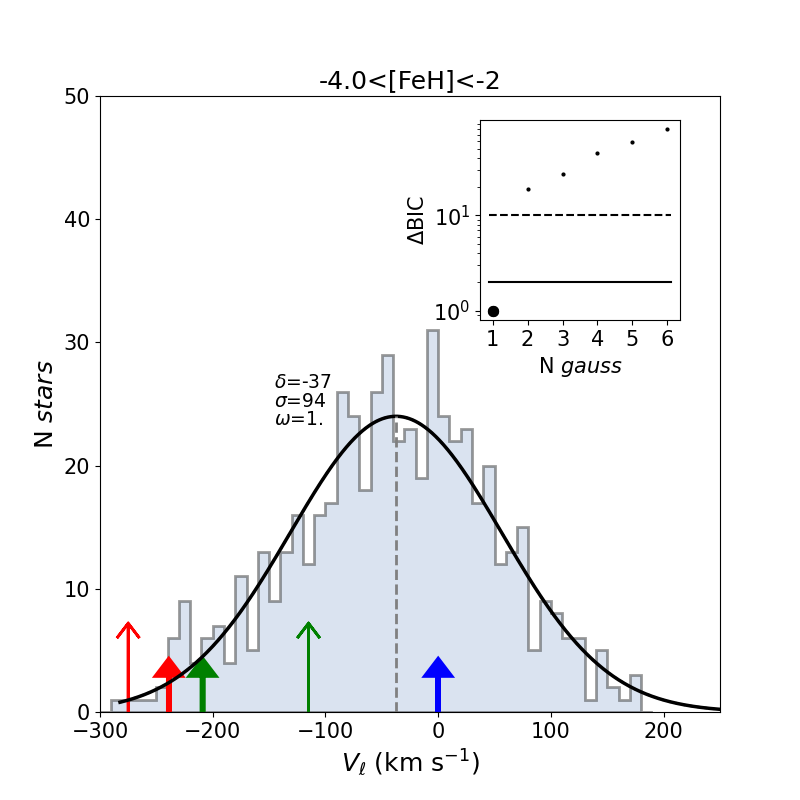} 
\includegraphics[scale=0.35]{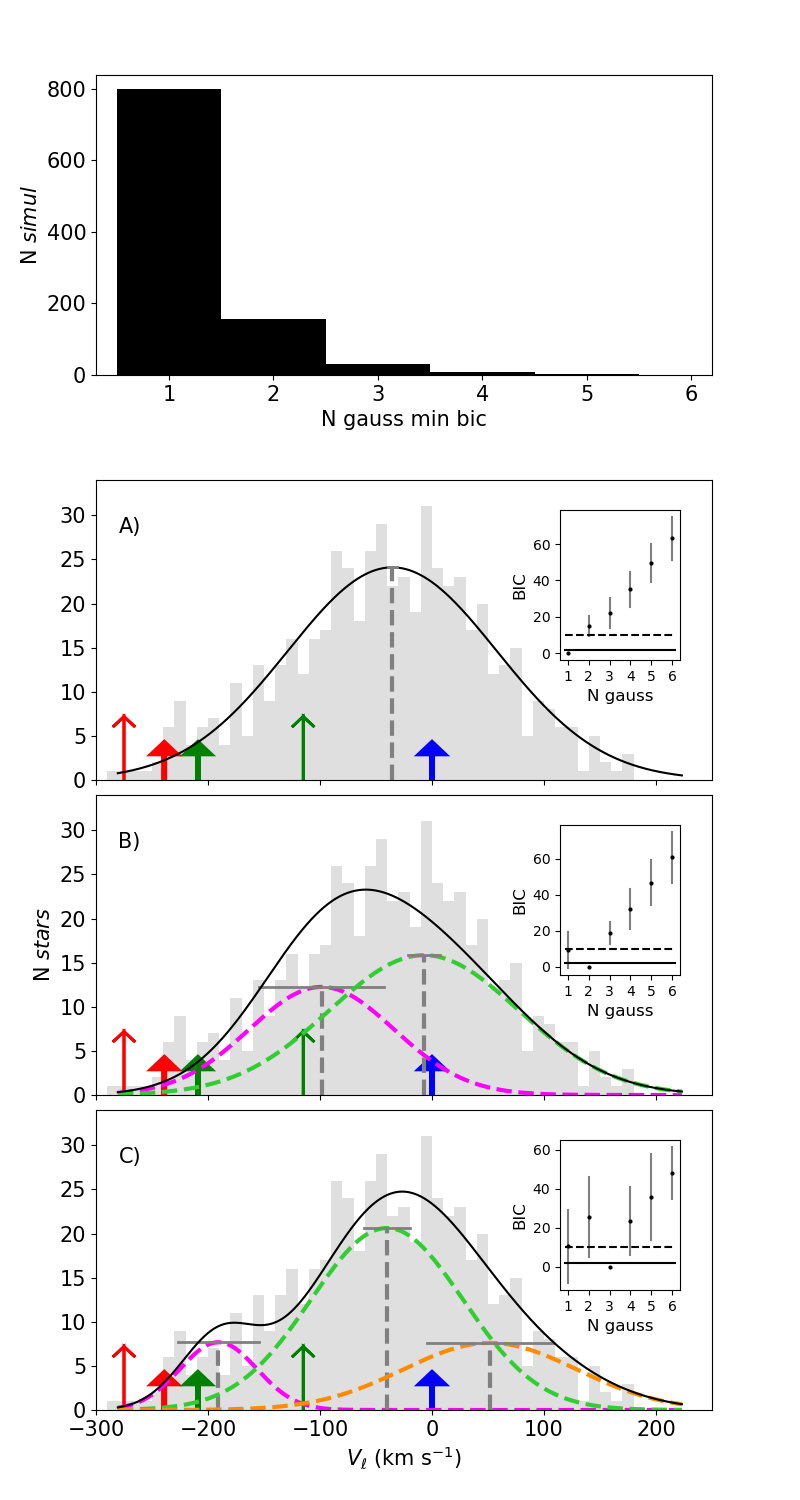}  
\caption{Same as Figure \ref{m2}. In this case, the individual gaussians were computed from the realisations with a best fit of one-gaussian model (panel A), a two-gaussian model (panel B) or a three gaussian model (panel C).}
\label{m4}
\end{figure*}

Our working sample covers Galactocentric distances R from 8 to $\sim 15$\,kpc, with a peak at $\mathrm{R} \sim 9$\,kpc and $\mathrm{z}\sim0.5$\,kpc, increasing in distance from the plane, z, as R increases, as shown in Figure \ref{Rz}.

The $V_{\phi}$ distribution is a good discriminant for Galactic stellar populations, whereas metallicity alone overlaps significantly between populations (eg. the Galactic thin and thick discs, or the Galactic thick disc and halo).
Thus, different Galactic populations should appear as distinct groups in the $V_{\phi}$ (or, in our case, $V_{\ell}$) vs. [Fe/H] plane. With our $V_{\ell}$ derived from the Gaia EDR3 proper motions (Equations 1 to 6) and the photometric [Fe/H] in hands, we then proceed  to examine the $V_{\ell}$ distribution of Milky-Way stars towards the \ac\ as a function of metallicity.

We will evaluate the $V_{\ell}$ distributions in the following metallicity bins: $\mathrm{[Fe/H]} > -0.8$\, (to include the region where we would expect to find the bulk of the thin disc), $-1.5 < \mathrm{[Fe/H]} < -0.8$\, (the metallicity range where, based on the comparison with the spectroscopic [Fe/H] estimates, we see a large systematic deviation in the error function), $-2 < \mathrm{[Fe/H]} < -1.5$\, (where the \textit{metal-weak} thick disc has been detected in previous work - \citealt{bessell85,ruchti11, kordo13mp} among others), $-4 < \mathrm{[Fe/H]} < -2$\, (the most metal-poor stars in our sample). The typical mean $V_{\ell}$ uncertainty, as derived from Gaia proper motions following the method presented in Section \ref{sec:vl_computation}, is of 3, 7, 22 and 28 \kms, respectively in each metallicity range.


\subsection{Examining the Galactic \ac\ on the $V_{\ell}$ vs. [Fe/H] plane}




Figure \ref{chachi} shows $V_{\ell}$ as a function of [Fe/H] for our sample. We recall that negative $V_{\ell}$ corresponds to prograde motion and positive $V_{\ell}$ to retrograde orbits. The contour lines, showing the 33, 66, 98, 99 and 99.9 \% of the cumulative distribution of the stars in this plane, highlight several substructures that we discuss below.


We see that the \ac\ is dominated by metal-rich stars, with [Fe/H] higher than $-1$\,. The [Fe/H] and $V_{\ell}$ distributions on the top and right panels of Figure \ref{chachi} quantitatively verify that our sample is dominated by a chemical and kinematical thin disc, with a metallicity distribution function (MDF) and a $V_{\ell}$ distribution that peak at $-0.5$ and $-225$\kms, respectively.  
We will however refrain from analyzing too deeply this population, since we know that our sample is biassed against the more metal-rich stars, that would be part of this thin disc.

It is worth noting the intermediate-metallicity population ($\sim-0.7$) extending from $V_{\ell} \sim-200$\kms to $V_{\ell} \sim+100$\kms, evidencing the presence of the thick disc and of the so-called \textit{Splash} component \citep{belokurov20}. The \textit{Splash}, detected from Gaia DR2 data -- see also \citet{babusiaux18}, \citet{haywood18}, \citet{fa19b} and \citet{dimatteo19} -- is thought to be a stellar population comprised of disc stars heated to halo-like kinematics likely because of the merger with the satellite galaxy Gaia-Enceladus-Sausage around 10 Gyrs ago. The continuous overdensity of stars at $\mathrm{[Fe/H]} > -1$\, extending towards increasing $V_{\ell}$ resembles such a disc-heated population.

One of the most puzzling stellar substructure unveiled by this analysis is the continuous sequence of stars moving with thin-disc-like $V_{\ell}$, not limited to the regime $\mathrm{[Fe/H]} > -1$\, but extending down to extremely low metallicity values, $\mathrm{[Fe/H]} < -3$\,. This rotating thin-disc-like sequence is clearly detectable down to $\mathrm{[Fe/H]} \sim -3$\,, as shown by the blue histogram on the top panel of Figure \ref{chachi}. 

The $V_{\ell}$ distribution of the stars at $\mathrm{[Fe/H]} < -1.5$\, is less conspicuous around thin/thick disc like velocities, i.e., $V_{\ell}\loa-150$\kms, in favour of a less rotating and hotter halo-like distribution. At those metallicities, stellar velocities are centered around $V_{\ell} \sim 0 $\kms, with a significant dispersion, of the order of $70-90$ \kms. Interestingly, the contour line which contains 99\% of the sample in Figure \ref{chachi} shows a slightly over dense area centered at a $V_{\ell}$ $\sim$ 0 \kms but with a narrower dispersion than the bulk of the halo stars, and it extends from the metallicities where the \textit{Splash} is detected, down to $\mathrm{[Fe/H]} \sim -2$\,. This hints at a possible link between the halo (at low metallicities), and the Splash (at higher metallicities), which will be further discussed in Sect. \ref{discuss_thick} -- see \citet{Kordopatis2020} with a similar conclusion. 

At low latitudes towards the \ac, the vertical velocity component, $V_{z}$ is approximately $V_{b}\cos(b)$ (see Equations \ref{ww} and \ref{vz}). We take advantage of this fact to evaluate how the different detected populations are characterized based on their vertical velocities. 
Figure \ref{Vb} shows the $V_{\ell}$ vs. [Fe/H] plane, colour-coded by $V_{b}$, split in bins of z. Before inspection, we recall that, because of our selection function, there is a correlation between z and R, i.e., stars at high z are also located at large R (see Figure \ref{Rz}). 
Close to the plane, at $\mathrm{z} < 0.5$\,kpc, the dominating population is the disc, with metallicities higher than $-1$, prograde orbits and low vertical velocities. In lower proportions, there are signatures of the halo and the \textit{Splash}. The number of retrograde stars, however, is very low and none with large $+V_{\ell}$. It is worth noting the presence of fast rotating very metal-poor stars down to $\mathrm{[Fe/H]} < -3$\, showing very low $V_{b}\cos(b)$, suggesting that these stars remain close to the plane. At $0.5 < \mathrm{z} < 1$\,kpc, where most of our sample is located, the fraction of non-rotating stars increases, and so does the fraction of retrograde stars. The thin-disc tail towards very low-metallicity values is also clearly detectable, with vertical velocities close to $V_{b}\cos(b)$ $\sim$ 0. 
As z increases, this very low-metallicity thin disc tends to disappear, and those stars with $\mathrm{[Fe/H]} < -2$\, moving in prograde orbits have also high vertical velocities, suggesting that they are the prograde tail of the velocity distribution of the halo. 

The presence of a disc and a halo population is clear in stars with metallicities from $-2$ to $-0.2$, up to $\mathrm{z} \sim 2$. At $\mathrm{z} > 2$\,kpc (and $\mathrm{R} > 11$\,kpc) stars show the more classical $V_{\ell}$ distribution of a kinematical halo centered at $0$\kms with a large dispersion at low metallicities and disc stars located at high metallicities, with no clear sign of very low-metallicity disc or high-metallicity \textit{Splash}. The \ac\,is close to the line of nodes of the Galactic warp \citep{lc02, yusifov04, momany06, reyle09,  chen19, skowron19} and, thus, the presence of disc stars at high z due to this feature is not expected. Besides, the amplitude of the warp it was always found lower than 2\,kpc at R $< 14$\,kpc -- see \citet{cheng2020} and references therein. Furthermore, it is not expected to have a gradient of more than a few \kms ($5-6$ between $8-14$\,kpc) in vertical velocities, and no variation in the rotational velocity  -- see figure 3 in \citet{poggio20}. Consequently, the effect of the warp on our spatial and kinematical distribution is negligible.

Figure \ref{Vb_histos} shows the $V_{\ell}$ distribution, now split in low- and high-metallicity stars, with $\mathrm{[Fe/H]} < -1.5$\, and $\mathrm{[Fe/H]} > -1.5$\,, respectively, and in high and low vertical velocities, with $|V_{b}\cos(b)| < 50$\,\kms and $|V_{b}\cos(b)| > 50$\,\kms, respectively. We choose 50\,\kms as the boundary to distinghish between disc and halo stars because this quantity is approximately $2\sigma$ the dispersion of the thin disc vertical motion, i.e., $\sigma = 25$\,\kms\, \citep{antoja21}. A vertical red dashed line marks the $V_{\ell}$ value lower than which stars move with thin-disc-like rotation. This Figure clearly shows that stars with $\mathrm{[Fe/H]} < -1.5$\, rotating like the thin-disc also have low vertical velocities, and there are almost no such kind of stars with high vertical motions.

\subsection{Contamination of metal-rich stars at low metallicities.}
\label{contamination}

Even though the comparison of photometric and spectroscopic estimates is overall very good, 
we expect, nevertheless, our low-metallicity sample to be contaminated by some higher metallicity stars that have been badly calibrated. 
Because metal-rich stars outnumber metal-poor stars, even a low percentage of failures can create a significant contamination at the lowest metallicity bins. It is therefore  crucial to estimate this contamination for the correct interpretation of the properties of very metal-poor stars.

In particular, it is important to verify whether the sequence of fast rotating metal-poor stars are not in fact, metal-rich contaminators with an underestimated photometric [Fe/H]. We recall that $V_{\ell}$ has a very small uncertainty, and therefore this disc-like sequence cannot be the result of a low accuracy in $V_{\ell}$ estimation. To investigate the impact of the photometric [Fe/H] uncertainties on this sequence, we model 
a thin-disc-like MDF and convolve it with the error function of Pristine's photometric metallicities. 

We decide to model the thin disc MDF as a single Gaussian, with a mean of $-0.15$ and a sigma of $0.22$. We obtain these values from \citet{hayden14} based on the (R,z) location of the stars within our volume. 
We derive the error function of the Pristine [Fe/H] metallicities from the differences between the photometric and spectroscopic [Fe/H] estimates, for stars in the range $-0.2 < \mathrm{[Fe/H]}_{\mathrm{spec}} < 0$\,. Then, we convolve the MDF by the error function, as shown in Fig. \ref{model_cont}.

We estimate the number of thin disc contaminants present in the metallicity range where no thin disc stars are expected, i.e., lower than $\sim -0.7$ \citep{bensby14} by scaling our toy-model to the total number of stars moving with thin disc like kinematics present in our sample. We obtain the total number of kinematical thin disc stars by fitting our whole \ac\ sample with a mixture of gaussians using a Gaussian Mixture Model \footnote{We make use of the python module sklearn.mixture \citep{scikit-learn}}. This model relies on a clustering technique to calculate the probability of a star to belong to a particular data cluster, for a pre-determined number of clusters and a convergence threshold. We obtain the number of gaussians that best account for the data evaluating the fits that result from considering Gaussian Mixture models comprised by 1 up to 6 clusters without any priors. The range of considered Gaussians is an educated guess based on Figure 8 (we expect that all of the dominant populations of the Galaxy are present in our sample). The best fit is chosen by minimising the Bayesian Information Criteria (BIC). Then, we infer the weight of the gaussian centered on the value closer to $-238$\kms ($V_{\phi}$ of the LSR). We find that 78\% of our \ac\ stars move with thin disc like $V_{\ell}$.

Based on our conservative toy-model\footnote{A more complex MDF, with an asymetric low metallicity tail has also been investivated, resulting  to fewer contaminators at low metallicities. The simpler MDF we therefore adopt, is justified as being a conservative estimate of the contamination.}, we calculate the number of contaminants in each of the metallicity bins. At metallicities $-2 < \mathrm{[Fe/H]} < -1.5$\,, and $-1.5 < \mathrm{[Fe/H]} < -0.8$\,, we find that the contamination of metal-rich stars due the [Fe/H] error function is of 4 and 4045, respectively. The number of stars in our sample with velocities compatible with the thin disc at these two metallicity ranges is 176 and 9316, respectively. It is expected, then, 
that less than 1\% of stars are metal-rich contaminants in the range $-2 < \mathrm{[Fe/H]} < -1.5$\,, and around 45\% between $-1.5 < \mathrm{[Fe/H]} < -0.8$\,. 
Therefore, we conclude that, at low metallicities, the population of stars rotating with thin disc $V_{\ell}$ cannot be explained solely by the contamination from higher metallicity stars. In the next section, we will evaluate the statistical significance of this scarce population.

\subsection{Statistical significance of the observed stellar substructures.}
\label{histos_sec}

We proceed to quantitatively evaluate how the $V_{\ell}$ distribution changes as a function of the metallicity and to interpret such distributions in terms of underlying stellar populations. For this purpose, we use Gaussian Mixture Models and evaluate models combining from 1 up to 6 gaussians, in order to derive the number of components that best fit the $V_{\ell}$ distributions in each metallicity range.

Previous works have shown that for the thin and thick disc the shape of the $V_{\phi}$ distribution is not Gaussian \citep[e.g.,][and references therein]{sharma13}. However, at first order it is a good approximation and simple to model. We will discuss our results in this context, keeping in mind the non-gaussianity of the true distributions.

The results are shown in Figures \ref{histos_mr}, \ref{m2} and \ref{m4}. We plot the gaussian mixture fit for which we obtained the minimum Bayesian Information Criteria (BIC), along with the corresponding individual gaussians in dotted lines and the position of their means in dashed vertical lines. In addition, we show with vertical arrows the typical $V_{\phi}$  values for the thin and thick discs and the halo (red, green and blue, respectively): $-238$\kms for the thin disc (our adopted LSR value), $-208$\kms, assuming a 30 \kms lag for the thick disc with respect to the thin disc \citep{recioblanco14}, and 0 \kms for the halo \citep{fermani13}. We also plot with thinner longer arrows the expected mean $V_{\phi}$ values for the thin and thick discs in each metallicity range, taking into account the correlation of $V_{\phi}$ with metallicity measured by \citet{kordopatis17} -- see also \citet{spagna10}, \citet{kordopatis11}, \citet{yslee11}, \citet{vardan13}, \citet{allende16}, \citet{ref19}, i.e., $-12\pm3$\,km\,s$^{-1}$\,dex$^{-1}$ for the thin disc and $+47\pm10$\,km\,s$^{-1}$\,dex$^{-1}$ for the thick disc. The underlying histograms depict the observed $V_{\ell}$ distributions, in bins of 10 \kms for each corresponding metallicity range.

\subsubsection{Stellar substructures at high metallicities,$\mathrm{[Fe/H]} > -1.5$\,: The thin disc, the thick disc and the Splash.}



The panels in Figure \ref{histos_mr} show the best gaussian mixture fit for the two most metal-rich ranges: $-0.8 < \mathrm{[Fe/H]} < -0.2$\, and $-1.5 < \mathrm{[Fe/H]} < -0.8$\,. The models that best fit the data are a mix of 5 and 4 gaussians, respectively. At the top right corner of each panel we plot the BIC differences between the other models and the best fit. BIC differences larger than 2 provide positive evidence that the model with the lower BIC is indeed the preferred fit, while differences larger than 10 imply strong and robust evidence \citep{raftery95}. The resulting BIC differences in our Gaussian Mixture models are overall higher than 10, confirming the robustness of the results. 

In order to interpret the resulting gaussian mixtures we compare the centers of the individual gaussians with the $V_{\ell}$ values at which we would expect to find the mean thin disc and thick disc distributions (red and green arrows, respectively, see beginning of Section \ref{histos_sec}). Looking at the most metal-rich bin, $-0.8 < \mathrm{[Fe/H]} < -0.2$\,, we verify that gaussian $a$ is centered at the mean thin disc $V_{\phi}$ value considering the correlation with [Fe/H]. The mean of gaussians $b$ and $c$ are close to the expected $V_{\phi}$ for the thick disc. The other two gaussians, $d$ and $e$, show a significant lag with respect to the thick disc, with mean $V_{\ell}$ values of $-172$ and $-107$\kms, respectively. Taking into account that the thin disc and thick disc $V_{\phi}$ distributions are not gaussian, gaussian $b$ is likely comprised of the lower rotation tail of the thin disc and a fraction of thick disc stars, and gaussian $d$ accounts for the lower rotation thick disc and some of the \textit{Splash}.

The same applies for the metallicity range $-1.5 < \mathrm{[Fe/H]} < -0.8$\,, with the fast rotating gaussian $a$ indicating the presence of the thin disc, gaussian $b$ the thick disc, and gaussian $c$ likely containing contribution from both the thick disc and the \textit{Splash}. Both the thick disc and the \textit{Splash} increase in proportion as the metallicity decreases.



The mean value of each gaussian shifts to lower rotation values (higher $V_{\ell}$) with decreasing metallicity. For the thick disc, this is in line with what found in \citet{kordopatis17}. However, the thin disc also decreases its rotation as metallicity decreases, contrary to what is measured at higher metallicities, $-0.5 < \mathrm{[Fe/H]}< 0.5$\, -- see \citet{kordopatis17}, \citet{ref19}. 


In order to assess the robustness of this result, we re-evaluate $V_{\ell}$ from 100 Monte-Carlo realisations of the parallax and proper motions, based on their associated uncertainties. This analysis confirms our results, i.e., that at high metallicities, $\mathrm{[Fe/H]} > -1.5$\,, the thin disc is the dominating population, but the $V_{\ell}$ distribution requires at least two additional components. The best gaussian mixture models that fit the data reveal, on the one hand, that the thick disc is clearly present in the \ac\ at such metallicity values and, on the other hand, that there is significant presence of a stellar component of much smaller weight and a very wide $V_{\ell}$ dispersion that describes well the characteristics of the \textit{Splash}, as shown by \citet{belokurov2020}. They found slightly lower mean rotation and dispersion values ($V_{\phi} = 25$\,\kms and $\sigma(V_{\phi}) = 54 \pm 11$\,\kms\,, but their analysis was performed at higher distances from the plane ($2 < |z| < 3$\,kpc), where only the stars heated to higher energies can reach. 

\subsubsection{Stellar substructures at low metallicities,$\mathrm{[Fe/H]} < -1.5$\,: the extremely metal-poor disc and the halo.}

Fitting the data at $-2 < \mathrm{[Fe/H]} < -1.5$\, returns as the optimal model a mix of two gaussians. Figure \ref{m2} shows that their means are centered on $V_{\ell}$ values close to: $a$) the LSR, and $b$ the halo. As done for the more metal-rich stars, we evaluate the robustness of these results by performing 1000 Monte-Carlo simulations.


The right top panel of Figure \ref{m2} shows the distribution of the number of components of the model with the minimum BIC for each of the 1000 Monte-Carlo simulations. For most of the realisations the best model is a mix of two gaussians, although there are some cases where it is a mix of three and even a mix of four. For the realisations for which the number of components of the best fit model is the same we calculated the mean center and $\sigma$ of each of the gaussian components and display the gaussians inferred from these means in the panels below. As in the previous plots, we also plot in red, green and blue arrows the expected rotational velocities of the thin, thick discs and the halo.


At the top-right corner of each plot we evaluate the significance of the best fit model relative to the other ones,  by displaying the mean BIC differences between the first and formers, as well as and their standard deviations as error bars.


We see from the BIC differences that when the best fit is composed by two Gaussians, then it is robustly the preferred model. For realisations for which the best fit is composed of three or four gaussians, the BIC differences with the model of two gaussians is close to 2, which implies that the three or four gaussian models are not significantly better than the two gaussian model even if they show the minimum BIC. Also notable is the fact that the mean of gaussian $b$ in the case of 3 gaussians, is very poorly defined, showing a huge dispersion. This gives strong support to the fact that, taking into account the uncertainties in our measurements, the two gaussian model is the best fit for the data.

Interestingly, the two gaussians that compose this model are: 1) one dominating the distribution, centered at low $V_{\ell}$ close to 0 and showing a large dispersion, that we associate to the halo; and 2) one centered at, approximately, the typical $V_{\ell}$ value of the LSR, that we refer to as the metal-poor rotating stars.

It is also remarkable that this second gaussian is always present even in the cases where the best fit is comprised of three or four gaussians. In these cases it is the gaussian associated to the kinematical halo that splits into several components (a more prograde thick-disc-like one and another more retrograde).


In summary, the evidence for the presence of a kinematical thin disc like component is significantly robust at metallicities $-2 < \mathrm{[Fe/H]} < -1.5$\,. We recall that the expected contamination of metal-rich stars at this metallicity range due to Pristine [Fe/H] error function is less than 1\%, based on our estimates (see Sect.~\ref{contamination}). Obviously, this metal-rich contamination cannot account for the kinematical thin disc component detected. Also noticeable is the fact that this component does not sharply disappear at $\mathrm{[Fe/H]} = -2$\, but extends at very low values, down to $\mathrm{[Fe/H]} \sim -3.5$\,. 

On the other hand, our stellar sample at such low metallicities is dominated by a prograde halo (see Figure \ref{m2}). This component could be comprised by the sum of other substructures, like a thick-disc and even a retrograde stellar population. However, the total number of stars in this component is too low to draw such conclusions with good statistical significance.



We perform the same exercise for stars at $-4 < \mathrm{[Fe/H]} < -2$\,, obtaining that the data is well fit with a single gaussian centered at $-37$\kms (see Figure \ref{m4}). Instead of a Monte-Carlo, due to the low number of stars in this metallicity range, we applied a boostrap technique, i.e. we implicitely consider that the uncertainties on $V_{\ell}$ are smaller than the impact of low-number statistics on the sample realisation. We in fact verified that the variance of the resulting gaussian mixture is higher from the bootstrap technique than from the Monte-Carlo on $V_{\ell}$ uncertainties. We perform 1000 realisations, selecting each time a random sample of the same size of our dataset at this metallicity range (141), and evaluating several Gaussian Mixture Models (comprising from 1 to 6 gaussian components as for the previous metallicity ranges) for each of the 1000 realisations. 

The bootstrapping technique confirms the previous result with 80\% of the realisations returning a single gaussian as the best fit. The other 20\% fit the distribution with two, three and even five gaussians. As for the previous metallicity range, when comparing the BIC differences of the best fit model with the other models, the single gaussian component is robustly the best fit when it has the minimum BIC. Noticeably, all the best fit models suggest that the component closest to the kinematical halo, i.e., with a mean $V_{\ell}$ close to 0 \kms, has always a net prograde rotation. In the cases where the best fit is comprised by two components or more, there is always an additional more prograde component, revealing that the $V_{\ell}$ distribution at these very low metallicities is shifted towards a net prograde rotation.

The extension of metal-poor rotating stars to the extremely metal-poor regime is not statistically identified as a separate gaussian component (independently of the number of components), even though we do see it as a small excess around $V_{\ell}\sim -230$\kms.



Finally, we present in Table \ref{table:1} the different best fit gaussians with their corresponding mean <$V_{\ell}$>, standard deviations $\sigma$ and relative fractions \% along with the Galactic populations with which we associate each of them.

\begin{table*}
\caption{Mean, standard deviation and relative fraction of the gaussian components that best fit the $V_{\ell}$ distributions of Pristine stars towards the \ac. We specify on the first column the Galactic component that we associate to each Gaussian based on how they compare to the literature values for the thin-/thick discs and the halo. Since the velocity distributions of the Galactic components are not gaussians, some of the resulting fitted gaussians include stars of the non-gaussian tail of two Galactic components. We indicate these cases by associating the corresponding gaussian to more than one component.}             
\label{table:1}      
\centering          
\begin{tabular}{ | l | c c c | c c c | c c c | c c c | }     
\hline\hline       
 & \multicolumn{3}{c|}{$-0.8 < \mathrm{[Fe/H]} < -0.2$\,} & \multicolumn{3}{c|}{$-1.5 < \mathrm{[Fe/H]} < -0.8$\,} & \multicolumn{3}{c|}{$-2.0 < \mathrm{[Fe/H]} < -1.5$\,} & \multicolumn{3}{c|}{$-4.0 < \mathrm{[Fe/H]} < -2.0$\,} \\ 
\hline   
                       & <$V_{\ell}$> & $\sigma$ & \% & <$V_{\ell}$> & $\sigma$ & \% & <$V_{\ell}$> & $\sigma$ & \% & <$V_{\ell}$> & $\sigma$ & \% \\ 
                                       
   &  \multicolumn{2}{c}{(\kms)} &  &  \multicolumn{2}{c}{(\kms)} & & \multicolumn{2}{c}{(\kms)} & &  \multicolumn{2}{c}{(\kms)} &  \\

\hline
   Thin disc                & $-$245 & 13 & 37    & $-$237 & 16 & 46   & $-$233 & 19 & 11    & & & \\  
   Thin disc + Thick disc   & $-$220 & 13 & 28    & $-$211 & 18 & 28   &      &    &         & & & \\
   Thick disc               & $-$201 & 17 & 23    & $-$166 & 30 & 14   &  &  &     & & & \\
   Thick disc + Halo/Splash & $-$172 & 29 & 10     &      &    &        &      &    &         & & & \\
   Halo/Splash              & $-$107 & 75 & 2    & $-$60 & 82 & 12    & $-$38  & 86 & 89    & $-$37 & 94 & 100 \\
\hline  
\hline                
\end{tabular}
\end{table*}

\section{Discussion}


In the present work, we offer an analysis of the distribution of kinematics along a wide metallicity range ($-3.5 < \mathrm{[Fe/H]} < -0.2$\,), from the analysis of the Pristine survey, a unique sample in terms of metallicity estimates (homogeneously derived), especially at the low metallicity end. In this Section we discuss the implications of our findings in the context of Galaxy formation and the novelties that they uncover compared to previous analyses found in the literature.


\subsection{Implications on the formation of the very and extremely metal-poor thin disc.}

Our findings strongly suggest the presence of a kinematical thin disc population down to $\mathrm{[Fe/H]} \sim -2$\,. This population, at low metallicities, rotates significantly faster than the present-day thick disc, with velocities closer to the thin disc. Indeed, the $V_{\ell}$ that we measure for this population are lower than the ones displayed by the thick disc detected at higher metallicites. There seems to be a continuous distribution of stars moving in thin disc like orbits from very low ($\mathrm{[Fe/H]} \sim-2$\,) up to the highest metallicities. There are fast-rotating stars at even lower metallicities, down to $\mathrm{[Fe/H]} \sim -3.5$\,, observed close to the plane, and with low vertical motions. However, we did not find statistical evidence that these stars belong to a separate stellar population, linked to the thin-disc-like one detected at higher metallicities, rather than the prograde rotating tail of the halo distribution centered at $V_{\ell}$, with a dispersion of around 100 km/s. In any case, it is stricking to see the asymmetric shape of the $V_{\ell}$ distribution for those stars at $\mathrm{[Fe/H]} > -2$\,, with more fast rotators in the prograde rather than in the retrograde sense.

Extremely ($\mathrm{[Fe/H]} < -3$\,) and even ultra ($\mathrm{[Fe/H]} < -4$\,) metal-poor stars moving in disc-like orbits were recently discovered by \citet{sestito19}, and, since then, several works have confirmed their existence \citep{sestito20, dimatteo19, dimatteo20, venn20, carter21, cordoni21}. From a full orbital characterization of all known ultra metal-poor stars with radial velocities \citet{sestito19} found a large fraction of them confined to $\mathrm{z}_{max}$ < 3\,kpc to the plane (26\%), and two of them (5\%) moving in prograde circular motions. \citet{sestito20} also discovered, among stars with $\mathrm{[Fe/H]}<-2$\, from the Pristine and LAMOST surveys, a significant number of them moving with thick disc like orbits. In addition, \citet{dimatteo20}, comparing stars from ESO's Large Program ``First Stars'' with other stellar samples, also discovered a subsample moving on thick-disc-like orbits, i.e., sharing the locus of the thick disc in the Toomre diagram. Comparing with other stellar samples covering all metallicity regimes from $\mathrm{[Fe/H]} < -4$\, up to $\mathrm{[Fe/H]} > -0.5$\, they showed that, at every metallicity range, there is a kinematical disc and a halo population coexisting. In the present work we confirm from an homogeneous sample that this is true, at least down to $\mathrm{[Fe/H]} \sim -2$\,, except that our kinematical disc rotates more like a thin disc than a thick disc. The fact that our sample are mostly confined at low z implies that we probe this thin disc population preferentially compared to other studies. 


Based on chemical evolution models that consider a disc formed in an inside-out fashion \citep{larson76, cole2000}, the outer regions form  stars at a slower rate, due to a delayed accretion of gas at these radii. This gas dilutes the Galactic interstellar medium and decreases its metallicity. Thus, subsequent stars formed in the outer regions are more metal-poor than stars born in the inner disc at the same epoch, and, in addition, by that time the gas there may have settled already into a disc configuration. Inside-out chemical evolution models such the two-infall model \citep{chiappini97,chiappini01} have been applied to explain the age-metallicity, age-[$\alpha$/Fe], the [$\alpha$/Fe] vs. [Fe/H] dichotomy or chemical abundance trends in Galactic stellar populations by several groups \citep{spitoni19,spitoni20,minchev13,minchev16,minchev17,grisoni19, grisoni20a,grisoni20b,grisoni20c}. Interestingly, \citet{spitoni21} reproduce a decrease of around 1 dex in metallicity at radius between 8 and 10\,kpc, and their predicted [Fe/H] distributions show a metal-poor tail down to at least $-2$\,. Our disc stars down to $\mathrm{[Fe/H]} \sim-2$\, would have formed, then, after this second infall of gas into the Galaxy. This scenario would explain their thin disc like motion. It is, however, difficult to explain the existence of stars at $\mathrm{[Fe/H]} < -2$\,. For these EMP stars the formation in the early disc seems more feasible. Further analysis of the chemical abundances of these stars is needed, for instance regarding the [$\alpha$/Fe] ratio, to clarify whether there is a link between fast rotating stars at metallicities higher and lower  than $\mathrm{[Fe/H]} \sim -2$\,.



Stars with metallicities $\mathrm{[Fe/H]} < -2$\, are expected to be formed during the first 2-3 Gyr after the Big Bang \citep{elbadry18} and at this epoch the Galaxy was still assembling through the hierarchical merger of smaller systems. Under this scenario, stars either accreted or born in-situ in a proto-disc were expected to be pressure-supported from the very beginning of their formation or heated into halo-like kinematics due to the impact of the mergers. Thus, finding stars at such low metallicities still moving in a disc configuration is puzzling. \citet{dimatteo19} pointed out that the earliest star formation in the Milky-Way took place in a thick disc configuration that was partially heated due to mergers, based on the observational results evaluating the impact of the merger with Gaia-Enceladus in the kinematics of the chemically defined thick disc (see figures 12 and 13 in \citealt{dimatteo19}). On the other hand, \citet{sestito21} explored NIHAO-UHD cosmological zoom in simulations \citep{buc20} and ended up with two possible and non-exclusive scenarios: 1) these UMP stars in planar orbits were accreted through minor mergers onto the Galactic plane and deposited in the disc by dynamical friction; and 2) they were born in gas-rich building blocks that led afterwards to the formation of the Milky-Way proto-disc.

The recent work of cosmological simulations by \citet{park20} suggested that spatially defined thick discs originate first as thin discs and then evolve to a slower rotating and spatially thicker configuration due to orbital difusion. The final thin disc observed at redshift zero is younger and rotates faster than the thick disc as a consequence of disc settling. In fact, their simulations (with the highest resolution obtained up to now) showed that both thin and thick discs are \textit{two parts of a single continuous disc component that evolves with time as a result of the continued star formation of thin disc stars and disc heating}. If our metal-poor thin disc was formed in the early disc, then these stars should have conserved their kinematics, not altered by the diffusion that formed the thick disc.

Further characterization is needed to clarify the origin of these stars and their chemical composition and ages would be very valuable. For example, the [$\alpha$/Fe] ratios could help to discriminate the different scenarios proposed. In the case where these stars formed in the early disc the [$\alpha$/Fe] would be high, similar to the one characterizing the thick disc at $\mathrm{[Fe/H]} \sim -1$\, ($\mathrm{[}\alpha\mathrm{/Fe]}\sim+0.4$\,). However, if they were formed later from interstellar medium diluted by a second infall of gas then they would present lower [$\alpha$/Fe] ratios, similar to the chemically metal-rich thin disc ($\mathrm{[}\alpha\mathrm{/Fe]}\sim<0.2$\,). If the latter is the true, then these stars could help to better constrain the intensity, initial time and duration of this second infall.



\subsection{The thick disc, the Splash and the halo.}
\label{discuss_thick}


The presence of the thick disc in the \ac\ is necessary to explain the observed rotational velocity distribution. This detected thick disc component also follows the correlation of the mean rotational velocity with metallicity inferred in previous works as the metallicity decreases ($47\pm10$\kms, e.g.  \citealt{kordopatis17}). Our analysis reveals the presence of the thick disc down to, at least, $\mathrm{[Fe/H]} \sim -1.5$\,. In our sample this thick disc component blurs kinematically into the halo when decreasing to metallicities lower than $-1.5$\,, but there are hints of its presence at $-2 < \mathrm{[Fe/H]} < -1.5$\,. These are the values at which the metal-poor (or metal-weak) thick disc has been detected in previous works: \citet{norris85}, \citet{morrison90}, \citet{ruchti11}, \citet{kordo13mp}, \citet{beers14}, \citet{carollo19}, but see subsequent discussions from \citet{twarog94}, \citet{twarog96}, \citet{chiba2000},
\citet{beers2002}, \citet{reddy08}, \citet{brown08}, 
\citet{kordo13mp}, \citet{hawkins15}, \citet{lizhao17},
\citet{hayes18}.

The fraction of the thick disc at $\mathrm{[Fe/H]} < -1$\, inferred in the previous works of \citet{anbeers2020} and \citet{dimatteo19} was around 40\%. We recover a lower fraction: the weight of our fitted gaussians with a mean close to the typical mean $V_{\phi}$ of the thick disc is 14\% in the range between $-1.5 < \mathrm{[Fe/H]} < -0.8$\, and 19\% at $-2 < \mathrm{[Fe/H]} <-1.5$\,.


The extension of metal-rich stars at $\mathrm{[Fe/H]} > -1$\, towards non-rotating and counter-rotating $V_{\ell}$ values is compatible with the kinematical effect of a heated disc. Recently, signatures of a heated thick disc (the so-called \textit{Splash}) have been detected in \citet{babusiaux18}, \citet{fa19b}, \citet{belokurov20}, \citet{dimatteo19}, \citet{gallart19}, and specifically in the \ac\ by \citet{anbeers20}, as a consequence of the merger with the dwarf satellite Gaia-Enceladus-Sausage \citep{belokurov18, helmi18, haywood18, dimatteo19}.  Besides, previous work \citep{dimatteo19} showed that the kinematical signature is more prominent at $-1 < \mathrm{[Fe/H]} < -0.3$\,, in line with our results. Recent analysis of cosmological simulations \citep{grand20} have found evidence of scatter of stars due to gas-rich mergers like GES leaving stars with a broad range of velocities that link the thick disc with the inner halo (although see also \citealt{amarante20} for an alternative origin for the \textit{Splash} due to clump scattering). On the other hand, there have also been  detections of a heated thin disc, such as the overdensities at the Monoceros and TriAnd \citep{bergemann18, hayes18triand, fa19b}, likely due to the interaction with Saggitarius \citep{thomas19, laporte19}. A further analysis of such heated metal-rich stars is needed to identify the several processes among our sample, which is beyond the scope of this paper.

The non-rotating and retrograde overdensity of stars extends down to the lowest metallicities. \citet{anbeers20} in their analysis of \ac\ stars detected a stellar overdensity distinct to GES, the \textit{Splash} and the thick disc, with a net prograde rotation. They called it the in-situ halo. Our results are in agreement with this finding. This component would be different than the metal-poor thick disc and a retrograde population, as shown by the different gaussians that fit the data. However, contrary to what was observed in \citet{anbeers20} at the low z range that we are studying, we see that this prograde halo dominates the distribution at such low-metallicities compared to the disc and retrograde components.



These results reinforce the power of photometric surveys such as Pristine. Right now, the Pristine metallicity estimates are linked to the SDSS photometry, There is, however, in preparation the conversion of the metallicity scale to Gaia photometry. This will hugely improve the Pristine analytical power. First, because there will be more stars with a metallicity estimate, not only stars with a SDSS photometric counterpart. And second, because Gaia photometry is more accurate than the SDSS one, and this will help to derive more precise metallicities.




\section{Conclusions}

We studied the rotational velocity distribution of Pristine stars towards the \ac. We took advantage of the fact that at this direction the rotational velocity component $V_{\phi}$ can be associated to the linear projection of the proper motion over the galactic longitude direction $V_{\ell}$. We estimated the rotational velocities from Gaia EDR3 proper motions and explored their distribution as a function of their metallicity. We inferred the metallicities from Pristine photometry, improving the calibration method by taking into account the stellar type of the stars. 

Our analysis revealed the presence of very metal-poor stars ($-3.5 < \mathrm{[Fe/H]} < -1.5$\,) moving with rotational velocities typical of the thin disc. In addition, although our sample is dominated by a kinematical thin disc at metallicities $\mathrm{[Fe/H]} > -0.8$\,, the presence of a thick disc component is also required to explain the observed velocity distribution. We also detect the signature of a heated disc at high metallicities, $\mathrm{[Fe/H]} > -1.5$\,. At $\mathrm{[Fe/H]} < -1.5$\,, our \ac\ sample is dominated by a kinematical halo with a net prograde rotation.

Upcoming large scale spectroscopic surveys such as WEAVE \citep{dalton16} and 4-MOST \citep{dejong19} at high- and low-resolution will significantly increase the number of stars in Pristine for which radial velocities will be measured. This will allow us to evaluate whether the rotational velocity distribution detected in this work shows the same pattern in other directions of the Galaxy and to fully characterize the kinematics with the vertical and radial velocity components. In addition, other chemical abundances will be measurable from spectroscopy. They will bring valuable information to constrain the scenarios that formed the observed galactic stellar populations.

\section*{Acknowledgements}
We thank Morgan Fouesneau for his valuable help providing a quality selection criteria to remove potential variable stars among Gaia data. We also gratefully thank Leticia Carigi for her insight in the interpretation of the present results based on chemical evolution models.
EFA, GK, VH, and NFM gratefully acknowledge support from the French National Research Agency (ANR) funded project ``Pristine'' (ANR-18-CE31-0017). We gratefully thank the CFHT staff for performing the Pristine observations in queue mode, for their reactivity in adapting the schedule, and for answering our questions during the data-reduction process. ES acknowledges funding through VIDI grant "Pushing Galactic Archaeology to its limits" (with project number VI.Vidi.193.093) which is funded by the Dutch Research Council (NWO). GT acknowledge support from the Agencia Estatal de Investigaci\'on (AEI) of the Ministerio de Ciencia e Innovaci\'on (MCINN) under grant with reference (FJC2018-037323-I). JIGH acknowledges financial support from the Spanish Ministry of Science and Innovation (MICINN) project AYA2017-86389-P, and also from the Spanish MICINN under 2013 Ram\'on y Cajal program RYC-2013-14875.
 
Based on observations obtained with MegaPrime/MegaCam, a joint project of CFHT and CEA/DAPNIA, at the Canada-France-Hawaii Telescope (CFHT) which is operated by the National Research Council (NRC) of Canada, the Institut National des Science de l'Univers of the Centre National de la Recherche Scientifique (CNRS) of France, and the University of Hawaii. The observations at the Canada-France-Hawaii Telescope were performed with care and respect from the summit of Maunakea which is a significant cultural and historic site.

This work has made use of data from the European Space Agency (ESA) mission
{\it Gaia} (\url{https://www.cosmos.esa.int/gaia}), processed by the {\it Gaia}
Data Processing and Analysis Consortium (DPAC,
\url{https://www.cosmos.esa.int/web/gaia/dpac/consortium}). Funding for the DPAC
has been provided by national institutions, in particular the institutions
participating in the {\it Gaia} Multilateral Agreement.

\section*{Data Availability}

The data underlying this article will be shared on reasonable request to the corresponding author.



\bibliographystyle{mnras}
\bibliography{vlfeh_mnras} 








\bsp	
\label{lastpage}
\end{document}